\documentclass[prd,twoside,twocolumn,superscriptaddress,nofootinbib]{revtex4-1}
\usepackage[colorlinks,citecolor=blue]{hyperref}
\usepackage{amsmath,amssymb}
\usepackage{graphicx}
\usepackage{xcolor}
\usepackage{multirow}
\usepackage{cancel}
\usepackage{color}
\usepackage{ulem}
\usepackage{float}
\usepackage{enumitem}

\newcommand{\beq}{\begin{equation}}
\newcommand{\eeq}{\end{equation}}
\newcommand{\bea}{\begin{eqnarray}}
\newcommand{\eea}{\end{eqnarray}}
\newcommand{\met}{\not{\!\rm E}_T}
\newcommand{\nn}{\nonumber}

\newcommand{\tabincell}[2]{\begin{tabular}{@{}#1@{}}#2\end{tabular}}

\allowdisplaybreaks

\begin{document}

\title{Searching for weak singlet charged scalar at lepton colliders}

\author{Qing-Hong Cao}
\email{qinghongcao@pku.edu.cn}
\affiliation{Department of Physics and State Key Laboratory of Nuclear Physics and Technology, Peking University, Beijing 100871, China}
\affiliation{Collaborative Innovation Center of Quantum Matter, Beijing 100871, China}
\affiliation{Center for High Energy Physics, Peking University, Beijing 100871, China}

\author{Gang Li}
\email{gangli@phys.ntu.edu.tw}
\affiliation{Department of Physics, National Taiwan University, Taipei 10617, Taiwan}

\author{Ke-Pan Xie}
\email{kpxie@snu.ac.kr}
\affiliation{Center for Theoretical Physics, Department of Physics and Astronomy, Seoul National University, Seoul 08826, Korea}

\author{Jue Zhang}
\email{juezhang87@pku.edu.cn}
\affiliation{Center for High Energy Physics, Peking University, Beijing 100871, China}

\begin{abstract}

A weak singlet charged scalar exists in many new physics models beyond the Standard Model. The discovery potential of the singlet charged scalar is explored at future lepton colliders, e.g. the CEPC, ILC-350 and ILC-500.  We demonstrate that one can discover the singlet charged scalar up to 118 GeV at the CEPC with an integrated luminosity of $5~\mathrm{ab}^{-1}$. At the ILC-350 and the  ILC-500 with an integrated luminosity of $1~\mathrm{ab}^{-1}$ such a discovery limit can be further improved to 136 GeV and 160 GeV, respectively. 

\end{abstract}

\maketitle

\section{Introduction}

The weak singlet charged scalar is an interesting signal of new physics (NP) beyond the Standard Model (SM) and often appears in NP models addressing on neutrino mass generation~\cite{Zee:1980ai,Aoki:2010ib,Krauss:2002px,Guella:2016dwo,Chekkal:2017eka}. The quantum number of the singlet charged scalar $S$ under the SM gauge group is $ (\mathbf{1},\mathbf{1})_{-1}$, where the first and second numbers inside the parenthesis indicate the quantum number of $SU(3)_{\mathrm{C}}$ and $SU(2)_{\mathrm{L}}$, respectively, and the subscript denotes the hypercharge. To describe the interactions between the singlet charged	 scalar $S$ and the SM fields, we adopt the effective field theory approach by writing down all possible operators that are invariant under the SM gauge group up to dimension-5. The renormalizable interactions of the singlet charged scalar $S$ with the SM leptons, i.e. $f_{\alpha\beta} \overline{\ell}_{\mathrm{L}\alpha} \ell_{\mathrm{L}\beta}^c S$, are severely constrained by current charged lepton rare decay data~\cite{McLaughlin:1999rr,TheMEG:2016wtm,Olive:2016xmw,Lavoura:2003xp}. We thus consider the dimension-5 operators built from the singlet charged scalar and the SM fields to describe the interaction of singlet charged scalar. We prove in our previous study~\cite{Cao:2017ffm} that, after performing field redefinitions and introducing gauge fixing terms, the bosonic operators do not contribute to the scalar decay at all. We end up with only four independent dimension-5 operators,
\begin{eqnarray}
\bar e_{\rm R} e^c_{\rm R} SS, \quad \overline{Q}_{\rm L} H u_{\rm R}S, \quad \overline{Q}_{\rm L}\widetilde{H} d_{\rm R} S^\dagger, \quad \overline{\ell}_{\rm L} \widetilde{H} e_{\rm R} S^\dagger.
\end{eqnarray}
Ignoring the three-body decay modes suppressed by phase space volume, we obtain the dominant decay modes of the singlet charged scalar $S^\pm$ (hereafter we use $S^\pm$ to represent the mass eigenstate of singlet charged scalar) as follows:
\begin{eqnarray}\label{eq:decay_channel}
S^- \rightarrow e^-\bar\nu, \quad \mu^-\bar\nu, \quad \tau^-\bar\nu, \quad q \bar{q}^\prime. 
\end{eqnarray}
In Ref.~\cite{Cao:2017ffm} we demonstrate that it is very promising to observe the singlet charged scalar at the Large Hadron Collider (LHC) as long as the scalar predominantly decays into a pair of leptons. On the other hand, the quark mode, as suffering from huge QCD background, cannot be detected in hadron collisions. 

In this work we consider both the Circular Electron Positron Collider (CEPC)~\cite{CEPC-SPPCStudyGroup:2015csa,CEPCStudyGroup:2018rmc} operating at a center-of-mass (c.m.) energy ($\sqrt{s}$) of $250~\text{GeV}$ and the International Linear Collider (ILC)~\cite{Djouadi:2007ik, BrauJames:2007aa} which is designed to run at $\sqrt{s}=350~\text{GeV}$ and $\sqrt{s}=500~\text{GeV}$, denoted as ILC-350 and ILC-500, respectively. In fact, the ILC also has a plan to operate at a energy of 250~GeV~\cite{Fujii:2017vwa}. We choose the CEPC as a benchmark lepton collider for $\sqrt{s}=250~{\rm GeV}$. Due to the limitations of the c.m. energies of the CEPC and ILC-350, we restrict our phenomenological study to the case of $m_S < m_t$, where $m_t$ denotes the top quark mass and $m_S$ denotes the singlet charged scalar mass. Although it is possible to search a singlet charged scalar heavier than top quark, i.e. $m_S > m_t$, at the ILC-500, we defer it for the future work due to the complexity arising from the top quark in final state.  Bearing in mind the fact that the Large Electron-Positron Collider (LEP) and the LHC constraints do not exclude a singlet charged scalar as light as 65~GeV~\cite{Cao:2017ffm}, we focus on a light charged scalar with $65~{\rm GeV}<m_S < m_t$  and do not distinguish the jet flavors in final state. When combining six different types of decay final state one is able to discover or exclude the singlet charged scalar up to a certain value of its mass,  \emph{irrespective} of its decay branching ratios. 

The rest of paper is organized as follows. In Sec.~\ref{sec:search} we investigate the discovery potential of the singlet charged scalar at the CPEC. The capability of the ILC operating at 350~GeV and 500~GeV is discussed in Sec.~\ref{sec:ilc}. Finally, we conclude in Sec.~\ref{sec:summary}. 

\section{Searching for $S^\pm$ at the CEPC} \label{sec:search}

\begin{figure}[b] 
\includegraphics[scale=0.5]{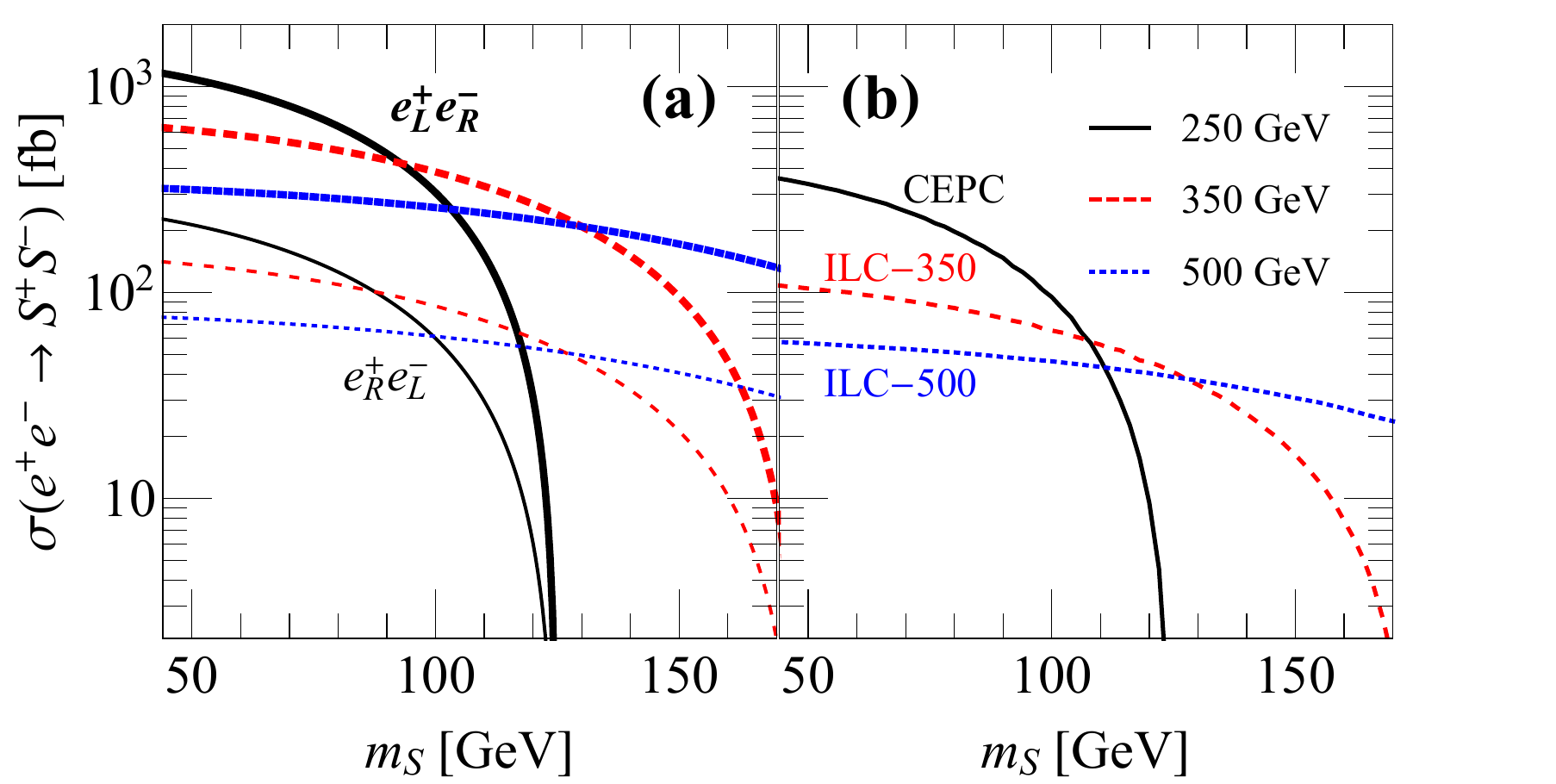}
\caption{Inclusive cross sections of the charged scalar pair production at lepton colliders with $\sqrt{s} = 250~\mathrm{GeV}$ (black solid), $350~\mathrm{GeV}$ (red dashed) and $500~\mathrm{GeV}$ (blue dotted). (a)~The initial electron and positron beams are polarized, i.e., $e^+_{R} e^-_L$ (thin) and $e^+_{L} e^-_R$ (thick); (b)~The polarizations of initial electron and positron beams are set to be those at the CEPC, ILC-350 and ILC-500. }
\label{fg:xsec_LC}
\end{figure}

In the electron-positron collision the singlet charged scalars are produced in pair through the channel of $e^+e^-\to \gamma/Z\to S^+S^-$. Note that there are dimension-5 operators which also can contribute to the production process, however, their contributions can be safely ignored.
The colliding electron and positron beams of the ILC are assumed to be $-80\%$ and $30\%$ right-handed polarized~\cite{Djouadi:2007ik,BrauJames:2007aa}, respectively, while the beams of the CEPC are unpolarized~\cite{CEPC-SPPCStudyGroup:2015csa,CEPCStudyGroup:2018rmc}.  Here for each beam we define its degree of polarization as $P_R - P_L$, where $P_R$ and $P_L$ are the fractions of right-handed and left-handed polarizations, respectively, satisfying $P_R + P_L = 1$. The cross sections of the scalar $S^\pm$ pair production with the polarized beams $e^+_R e^-_L$ and $e^+_L e^-_R$ are given by
\begin{eqnarray}
\sigma(e^+_R e^-_L) &=& \frac{2\pi \alpha^2}{3s^2} \left[ 1- \frac{(1-t_W^2)s}{2(s-m_Z^2)} \right]^2 \left(1- \frac{4 m_S^2}{s} \right)^{3/2}~, \nonumber \\
\sigma(e^+_L e^-_R) &=& \frac{2\pi \alpha^2}{3s^2} \left[ 1+ \frac{s_W^2s}{s-m_Z^2} \right]^2 \left(1- \frac{4 m_S^2}{s} \right)^{3/2}~,
\end{eqnarray}
where $s$ is the square of center-of-mass energy, $\alpha$ is the electromagnetic fine structure constant, $t_W^2 = \tan^2\theta_W$ and $s_W^2 = \sin^2\theta_W$ with $\theta_W$ being the Weinberg angle, and $m_Z$ is the $Z$-boson mass. Other polarization configurations of electron and positron beams $e_L^+e_L^-$ and $e_R^+e_R^-$ yield negligible contributions~\cite{MoortgatPick:2005cw}. As a result, we obtain the production cross sections the scalar $S^\pm$ pair at the CEPC and ILC-350 (500) as follows,
\begin{align}
&\sigma_{S^+S^-}^{\mathrm{CEPC}} = \frac{1}{4}\Big[\sigma(e^+_R e^-_L) + \sigma(e^+_L e^-_R)\Big], 
\\
&\sigma_{S^+S^-}^{\mathrm{ILC}} = 0.585 \times \sigma(e^+_R e^-_L)
+~0.035 \times \sigma(e^+_L e^-_R) .
\end{align}
In Fig.~\ref{fg:xsec_LC} we show the inclusive cross sections of the scalar $S^\pm$ pair production as functions of $m_S$ for $\sqrt{s} = 250~\mathrm{GeV}$ (black solid), $350~\mathrm{GeV}$ (red dashed) and $500~\mathrm{GeV}$ (blue dotted); the scenarios with fully polarized beams are shown in Fig.~\ref{fg:xsec_LC}(a), while in Fig.~\ref{fg:xsec_LC}(b) the beam polarizations are set to be those in the CEPC, ILC-350 and ILC-500. It can be seen that the cross section decreases dramatically in the region of $m_S\sim \sqrt{s}/2$. Therefore, the CEPC can cover only the light scalar region ($m_S\lesssim 100~{\rm GeV}$) while the ILC-500 could probe the heavy charged scalar region ($m_S\lesssim 160~{\rm GeV}$).    

\begin{figure}
\includegraphics[scale=0.5]{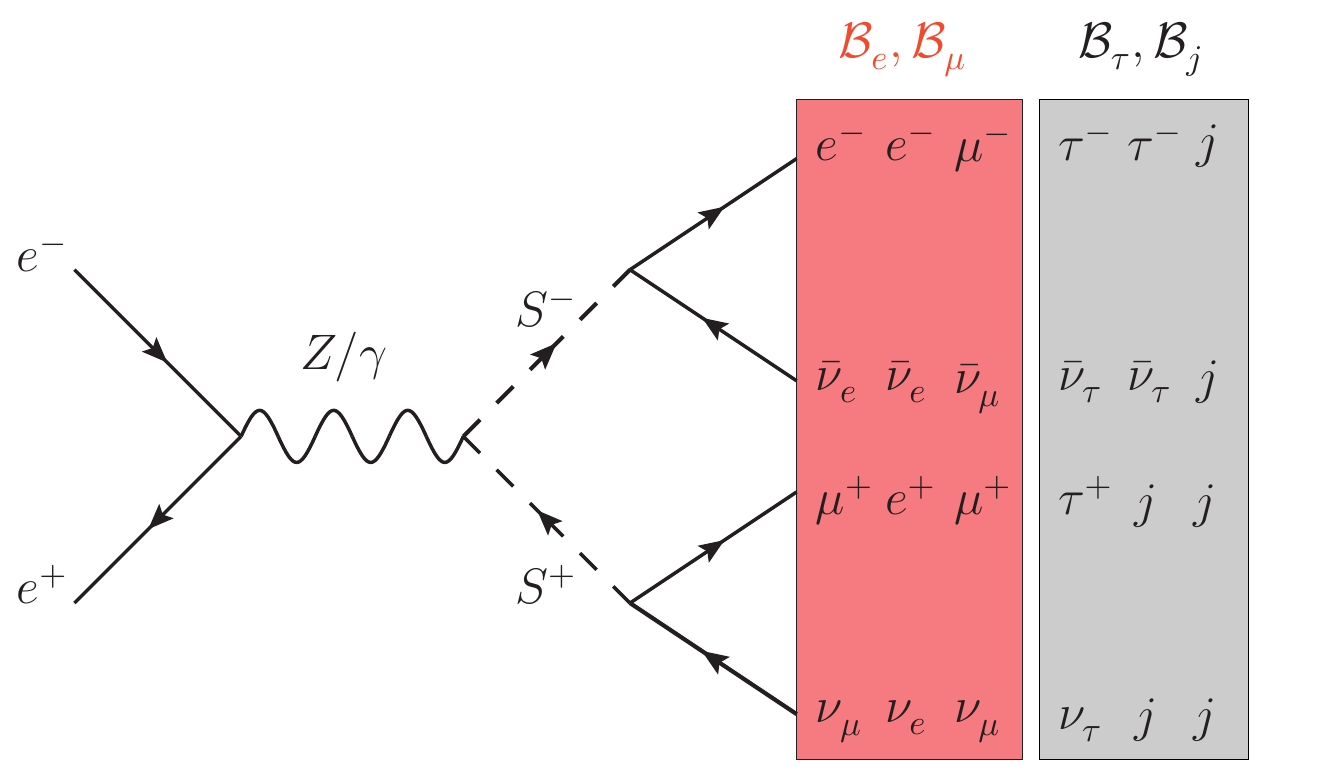}
\caption{The event topologies of singlet charged scalar pair production at the $e^+_{} e^-_{}$ colliders: the three dilepton channels shown in the red box are employed to constrain the decay branching ratios of $\mathcal{B}_e$ and $\mathcal{B}_\mu$, while $\mathcal{B}_\tau$ and $\mathcal{B}_j$ are probed by the decay channels listed in the gray box. For all decay channels charge conjugated processes are implied.}
\label{fig:sdecays}
\end{figure}

Figure~\ref{fig:sdecays} shows the six event topologies of singlet charged scalar pairs as follows:
\begin{align} \label{eq:channel}
&e^\pm\mu^\mp\nu\bar\nu, &&e^+ e^-\nu\bar\nu, && \mu^+\mu^-\nu\bar\nu, \nn\\
&\tau^+\tau^-\nu\bar\nu, &&\tau^\pm \nu j j,&& j j j j.
\end{align}
Fortunately, all the above six channels can be probed at the lepton colliders. We divide the six event topologies into two categories: one only consists of purely leptons, the other involves jets in the final state. The former is named as ``leptonic" mode while the latter as ``hadronic" mode. Both leptonic and hadronic decays of the tau lepton are considered in the study. We introduce four branching ratios ($\mathcal{B}_e$, $\mathcal{B}_\mu$, $\mathcal{B}_\tau$ and $\mathcal{B}_j$) to describe the $S^\pm$ decay, where $\mathcal{B}_i$ denotes the decay branching ratio of $S^\pm$ into the mode $i$. A detailed collider simulation is performed to explore the potential of probing the scalar $S^\pm$ at the CEPC and ILC. Our study shows that one can discover or exclude the charged scalar $S^\pm$ \emph{irrespective} of its decay branching ratios. 

As for the event generation, we mostly use {\sc MadGraph5\_aMC@NLO}~\cite{Alwall:2014hca} to generate events at parton level, unless in some cases, where the initial state radiation (ISR) and beamstrahlung effects are found to be significant, we switch to {\sc Whizard}~\cite{Kilian:2007gr,Moretti:2001zz}. We generate the universal {\sc FeynRules} output (UFO) model file of the singlet charged $S^{\pm}$ with {\sc FeynRules}~\cite{Alloul:2013bka} and implement it into {\sc MadGraph5\_aMC@NLO}. The packages {\sc Pythia 6}~\cite{Sjostrand:2006za} and {\sc Delphes}~\cite{deFavereau:2013fsa} are adopted for parton shower and detector simulation, respectively. For detector simulation we use the default cards shipped with {\sc Delphes}, i.e., the {\tt delphes\_card\_CEPC.tcl} and \texttt{delphes\_card\_ILD.tcl} cards for CEPC and ILC, respectively. In both cards we choose the minimal reconstructed transverse momentum ($p_T$) of jet to be $5~\mathrm{GeV}$~\cite{CEPC-SPPCStudyGroup:2015csa,CEPCStudyGroup:2018rmc} and the $\tau$-tagging efficiency as $60\%$~\cite{Djouadi:2007ik,BrauJames:2007aa}. 

Based on the event topologies of each decay mode, we consider several background processes. The dominant background processes are given as follows. The first kind of background involved is from the $\tau^+\tau^-$ production and $t\bar{t}$ pair production. The $t\bar{t}$ contributes at the ILC-350 and ILC-500. The $\tau$ leptons further decay into leptons to mimic the signal topologies of $e^\pm\mu^\mp \nu\bar{\nu}$, $e^+e^-\nu\bar{\nu}$ and $\mu^+\mu^-\nu\bar{\nu}$. We use {\sc Whizard} to generate this type of background to account for large ISR effects.  

\begin{figure}
\includegraphics[scale=0.35]{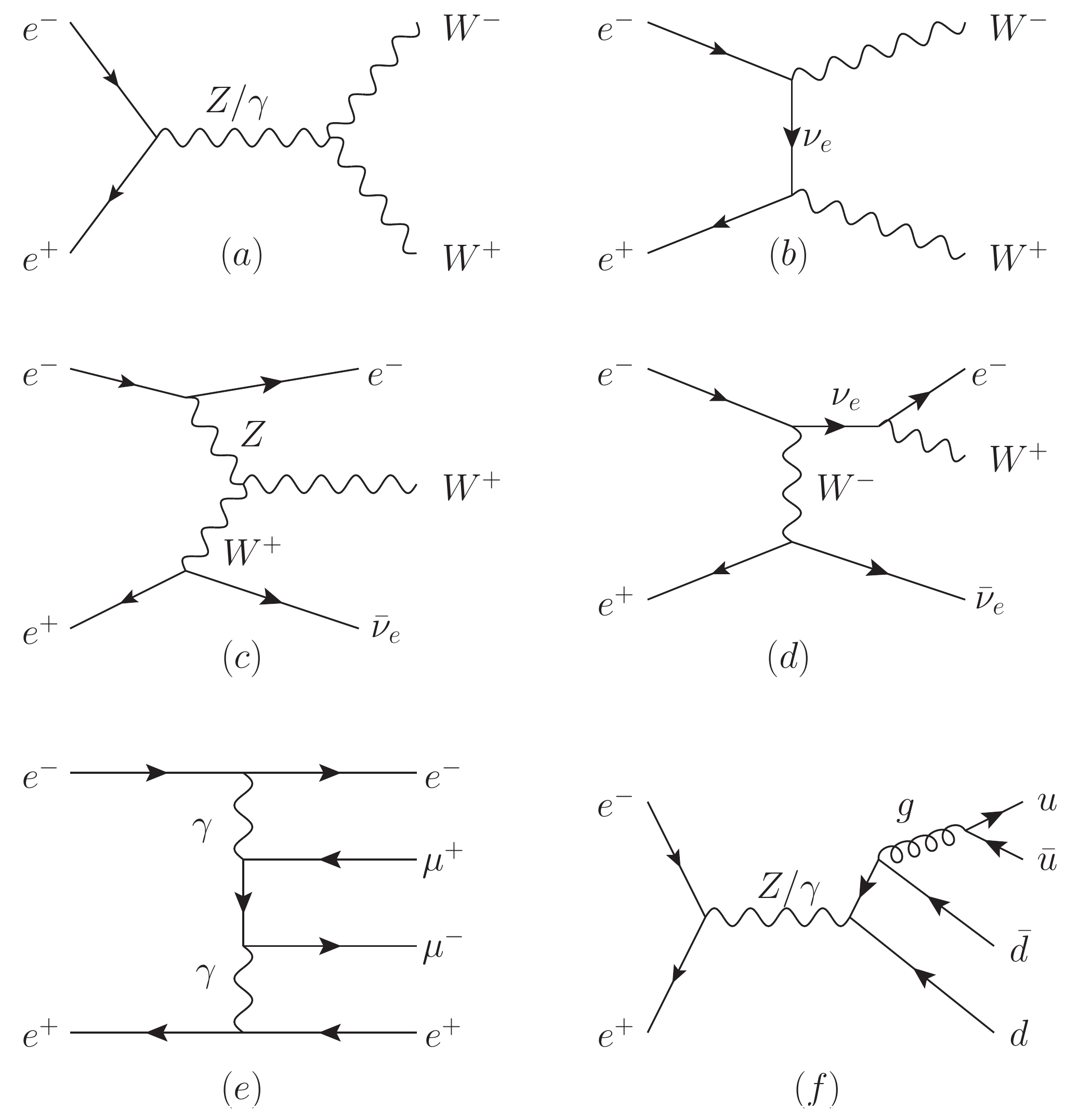}
\caption{Representative Feynman diagrams of four-fermion backgrounds at the $e^+_{} e^-_{}$ collider: (a, b) double-resonance backgrounds; (c, d) single-resonance backgrounds; (e, f) zero-resonance backgrounds including $e^+e^-\to e^+e^-\ell^+\ell^-$ and ``$jjjj$-QCD" processes. }
\label{fig:feyn}
\end{figure}

The second major backgrounds involve four fermions in the final state. The four fermions are produced through weak bosons ($W/Z$) or Higgs bosons ($h$) in the intermediate states. Experimental collaborations usually generate this type of backgrounds by directly considering $2\rightarrow 4$ processes~\cite{Mo:2015mza}. In order to identify the dominant backgrounds, we separate the backgrounds of four fermions into three sub-categories, depending on whether or not the intermediate $W/Z/h$ bosons are on mass shell. Figure~\ref{fig:feyn} displays the representative Feynman diagrams of the three subcategories:
\begin{enumerate}[leftmargin=*]
\item Double-resonance background : \\
it contains the production of two on-shell bosons, e.g. $e^+e^-\to W^+ W^-/Z Z/Zh$, where the Higgs boson $h$ further decays into a pair of fermions. See Figs.~\ref{fig:feyn}(a) and \ref{fig:feyn}(b). Due to the similarity in the decay channels of $W^\pm$ and $S^\pm$, and the closeness of their masses, the $W^+ W^-$ channel turns out to be the most important background. 

\item Single-resonance background: \\
it consists of only one on-shell boson in the intermediate state, e.g.   
$e^+e^-\to W^\pm\ell^\mp\nu$ ($\ell = e, \mu$), $W^\pm\tau^\mp\nu$, $W^\pm q\bar q^\prime$, $Z\ell^+\ell^-$, $Z\tau^+\tau^-$, $Z\nu\bar\nu$ and $Zq\bar{q}$, where the fermion pairs (such as $\ell^\pm\nu$, $\tau^\pm\nu$ and $q\bar q'$) are kept away from the $W/Z/h$ resonances. See Figs.~\ref{fig:feyn}(c) and \ref{fig:feyn}(d). Note that the ``single-resonance" processes also contain the diagrams listed in the first row when one of the electroweak bosons decays off-shell. 

\item Zero-resonance background: \\
it contains two major contributions. See Figs.~\ref{fig:feyn}(e) and \ref{fig:feyn}(f). The first one arises from the vector-boson-scattering (VBS) diagrams such as $e^+e^-\to e^+e^-\ell^+\ell^-$, while for the second type we include the gluon induced processes from the so-called ``reindeers" diagrams~\cite{Bardin:1994sc}. For this QCD induced type of background, we start generating events from the $jjj$ final state and then match them to the $jjjj$ final state. Such a matched QCD background sample is named as ``$jjjj$-QCD" hereafter. 
\end{enumerate}
The third type of backgrounds are $\gamma \gamma \rightarrow \tau^+ \tau^-$ with initial state photons from beamstrahlung. We simulate it with {\sc Whizard}.

Note that the triple gauge boson production (e.g. $e^+e^-\to W^+W^-Z$) and $e^+ e^- W^+ W^-$ can also mimic the signal on condition that two charged leptons or jets in the final state are not detected.  Such reducible backgrounds are not allowed by kinematics at the CEPC and are found to be less important at the ILC. We ignore them hereafter. 

\subsection{The $\ell^\pm\ell^{\prime\mp} +\met$ mode}

We begin with the $\ell^\pm \ell^{\prime\mp}\nu\bar{\nu}$ mode where $\ell$ denotes the electron or muon lepton. The signal events are characterized by two high-energy charged leptons and large missing energy ($\met$) from two unobserved neutrinos. As the $\tau^+\tau^-\nu\bar{\nu}$ channel can generate the signature of $\ell^\pm\ell^\mp +\met$ when the two tau leptons decay leptonically,  we consider the contribution of $\tau^+\tau^-\nu\bar{\nu}$ in the analysis and demonstrate that its contribution is small in comparison with the $e^\pm\mu^\mp \nu\bar{\nu}$ channels. For simplicity, we consider three benchmark models in our analysis of $\ell^\pm\ell^{\prime\mp}+\met$ mode as follows:
\bea
{\rm (A)} &:&  m_S = ~70~\mathrm{GeV},~\mathcal{B}_e = \mathcal{B}_\mu = 0.5,\nn\\
{\rm (B)} &:&  m_S = 100~\mathrm{GeV},~\mathcal{B}_e = \mathcal{B}_\mu = 0.5,\nn\\
{\rm (C)} &:&  m_S = 100~\mathrm{GeV},~\mathcal{B}_\tau = 1.\nn
\eea
Table~\ref{tb:dilepton_DF_CEPC} displays the inclusive cross sections (in the unit of femtobarn) of the signal and background events in the second column (denoted as ``no cut"). Note that the branching ratios of $S^\pm$ and $W^\pm/Z$ are not included. At this level the SM backgrounds dominate over the signal, yielding $\sigma_S/\sigma_B < 0.5\%$.

Since the SM backgrounds are sensitive to the lepton flavors, we distinguish the flavors of the two charged leptons in our analysis. We separate the signal events into two classes: one has different flavor (DF) leptons ($e^\pm\mu^\mp+\met$), and the other consists of same flavor (SF) leptons ($e^+e^-+\met$ and $\mu^+\mu^-+\met$). Treating the lepton flavors differently helps us with identifying the major background so as to introduce additional optimal cuts to suppress them. For example, a pair of SF leptons might arise from a on-shell $Z$ boson decay but a pair of DF leptons obviously cannot.  

\subsubsection{The DF case: $e^\pm\mu^\mp + \met $}
First consider the case of DF leptons. To compare the relevant background event rates to the signal event rate, we shall assume the integrated luminosity of the CEPC to be $100~{\rm fb}^{-1}$ and $5~{\rm ab}^{-1}$, and require both signal and background events to pass a set of selection cuts in event generation:
\beq \label{eq:selection_cuts}
n^{\ell}=2,\quad p_T^{\ell}> 5~\mathrm{GeV},\quad|\eta ^{\ell}| < 3, \quad \met>5~{\rm GeV},
\eeq
where $n^{\ell}$ denotes the number of charged leptons $\ell^\pm$ ($\ell=e,~\mu$), $p_T^\ell$ and $\eta^\ell$ represent the transverse momentum and the rapidity of $\ell^\pm$, respectively. We further demand that the two charged leptons exhibit different flavors. The cross sections of the signals and the dominant backgrounds after the selection cuts are shown in the third column of Table ~\ref{tb:dilepton_DF_CEPC}, where the branching ratios of $S^\pm$, $W^\pm$ and $Z$ bosons are included. In the benchmark models (A) and (B), the charged scalars directly decay into the $e^\pm\nu$ and $\mu^\pm\nu$ modes equally, thus leading to a production rate of $e^\pm\mu^\mp \nu\bar{\nu}$ as 118.3~fb for $m_{S}=70~{\rm GeV}$ and 45.2~fb for $m_S=100~{\rm GeV}$ before the selection cut. We find that about 75\% of signal events pass the selection cuts while only about 50\% of those major backgrounds ($W^+W^-$ and $W^{\pm} \ell^{\mp}\nu$) survive. The difference can be easily understood from spin correlation effects. The charged scalar pairs, which are produced through $e^+ e^- \to Z/\gamma \to S^+S^-$, exhibit a $p$-wave angular distribution proportional to $d^{1}_{1,0}(\theta_S)=\sin\theta_{S}/\sqrt{2}$, where $\theta_S$ is the polar angle of $S^\pm$ momentum with respect to the beamline direction. Hence, the $S^\pm$ tends to have a large $p_T(\propto\sin\theta_S)$ and central rapidity,  thus more often passing the selection cut. That dramatically enhances the ratio of the signal to the total background; for example, $\sigma_{S}/\sigma_{B}\simeq 16\%$ for the model (A) while $\sigma_{S}/\sigma_{B}\simeq 6\%$ for the model (B).

\begin{table}
\caption{Cross sections (in the unit of femtobarn) of the signal and background events in the $e^\pm \mu^\mp+\met$ mode at the CEPC. The kinematic cuts listed in each column are applied sequentially. The last column shows the significance of discovery potential with an integrated luminosity of $1~{\rm fb}^{-1}$.}
\label{tb:dilepton_DF_CEPC}
\centering
\begin{tabular}{ c|c | c | c | c | c |c}
\hline
\multicolumn{2}{c|}{$e^\pm\mu^\mp+\met$ }& No cut & Selection & $M_{T2}$& $\cos\theta_{\ell^\pm}$& $\dfrac{\sigma_{S}}{\sqrt{\sigma_B}}\sqrt{\dfrac{\mathcal{L}}{{\rm fb}}} $\\
\hline
(A) & \tabincell{c}{$\sigma_{\rm S}$\\ $\sigma_{S}/\sigma_{B}$} & \tabincell{c}{236.6\\0.5\%} & \tabincell{c}{91.7\\ 15.7\%} & \tabincell{c}{64.5\\ 19.9\%} & \tabincell{c}{33.8\\66.1\%}& 4.73\\
\hline
(B) &\tabincell{c}{$\sigma_{\rm S}$\\ $\sigma_{S}/\sigma_{B}$}  & \tabincell{c}{90.3\\0.2\%}& \tabincell{c}{34.9\\ 6.0\%} & \tabincell{c}{29.8\\ 9.2\%} & \tabincell{c}{13.9\\ 27.2\%} & 1.94	 \\
\hline
(C)& \tabincell{c}{$\sigma_{\rm S}$\\ $\sigma_{S}/\sigma_{B}$} &\tabincell{c}{90.3\\ 0.2\%}	&\tabincell{c}{2.9\\ 0.5\%}	&\tabincell{c}{1.4\\ 0.4\%}	& \tabincell{c}{0.6\\ 1.2\%}& 0.08 \\
\hline
\hline
\multicolumn{2}{c|}{$W^+W^-$ }&16520	&	390.3	&	282.0	&	48.1 & -\\
\hline
\multicolumn{2}{c|}{$ZZ$ }& 1100	&	0.5	&	0.2	&	0.1 & -\\
 \hline
\multicolumn{2}{c|}{$W^\pm e^\mp \nu$ }&  906	&	52.9	&	37.6	&	2.1 &- \\
 \hline
\multicolumn{2}{c|}{$W^\pm\mu^\mp \nu$ }& 51.0	&	5.0	&	3.9	&	0.7 &-  \\
 \hline
\multicolumn{2}{c|}{$Z\ell^+\ell^-$ } & 527	&	1.2	&	0.1	&	0.1 & -\\
 \hline
\multicolumn{2}{c|}{$e^+e^-\ell^+\ell^-$ (VBS) } & 22740	&	30.1	&	0	&	0 &- \\
 \hline
\multicolumn{2}{c|}{$e^+e^-\tau^+\tau^-$ (VBS) } & 5038	&	12.0	&	0.3	&	0&- \\
 \hline
\multicolumn{2}{c|}{$\tau^+\tau^-$ } &4321	&	97.7	&	0	&	0 & - \\
\hline
\end{tabular}
\end{table}

In the model (C) the charged scalar is assumed to decay entirely into the $\tau^\pm\nu$ mode. Owing to the branching ratio of tau lepton decaying into electron or muon leptons, $\mathcal{B}(\tau^+\to e^+\nu\bar\nu)=\mathcal{B}(\tau^+\to \mu^+\nu\bar\nu)=17\%$, the signal rate in the model (C) is about ten times less than the rate in the model (B). Therefore, our analysis of the $\ell^\pm\ell^{\prime\mp}+\met$ mode is sensitive to $\mathcal{B}_e$ and $\mathcal{B}_\mu$.

\begin{figure}
\centering
\includegraphics[scale=0.32]{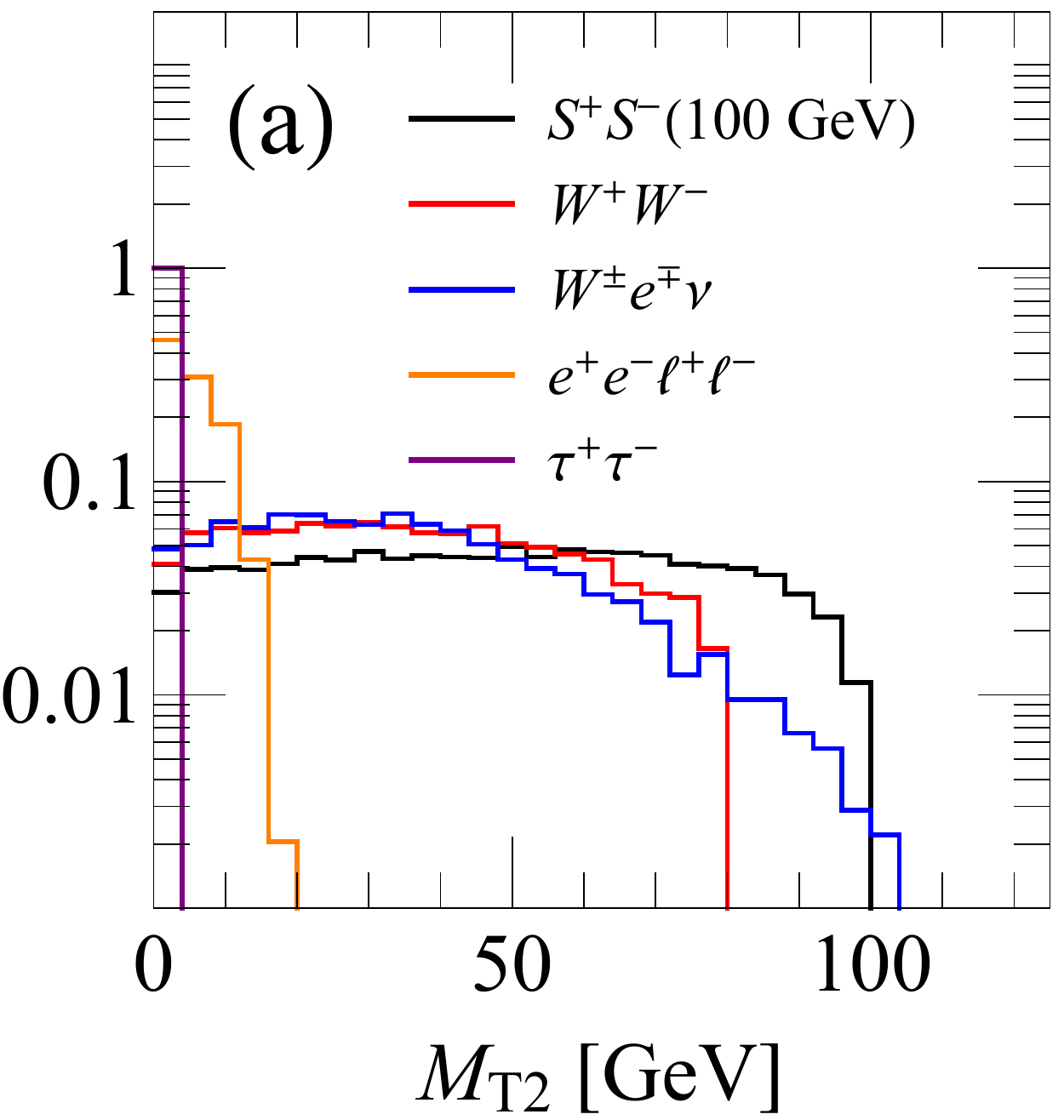}
\includegraphics[scale=0.32]{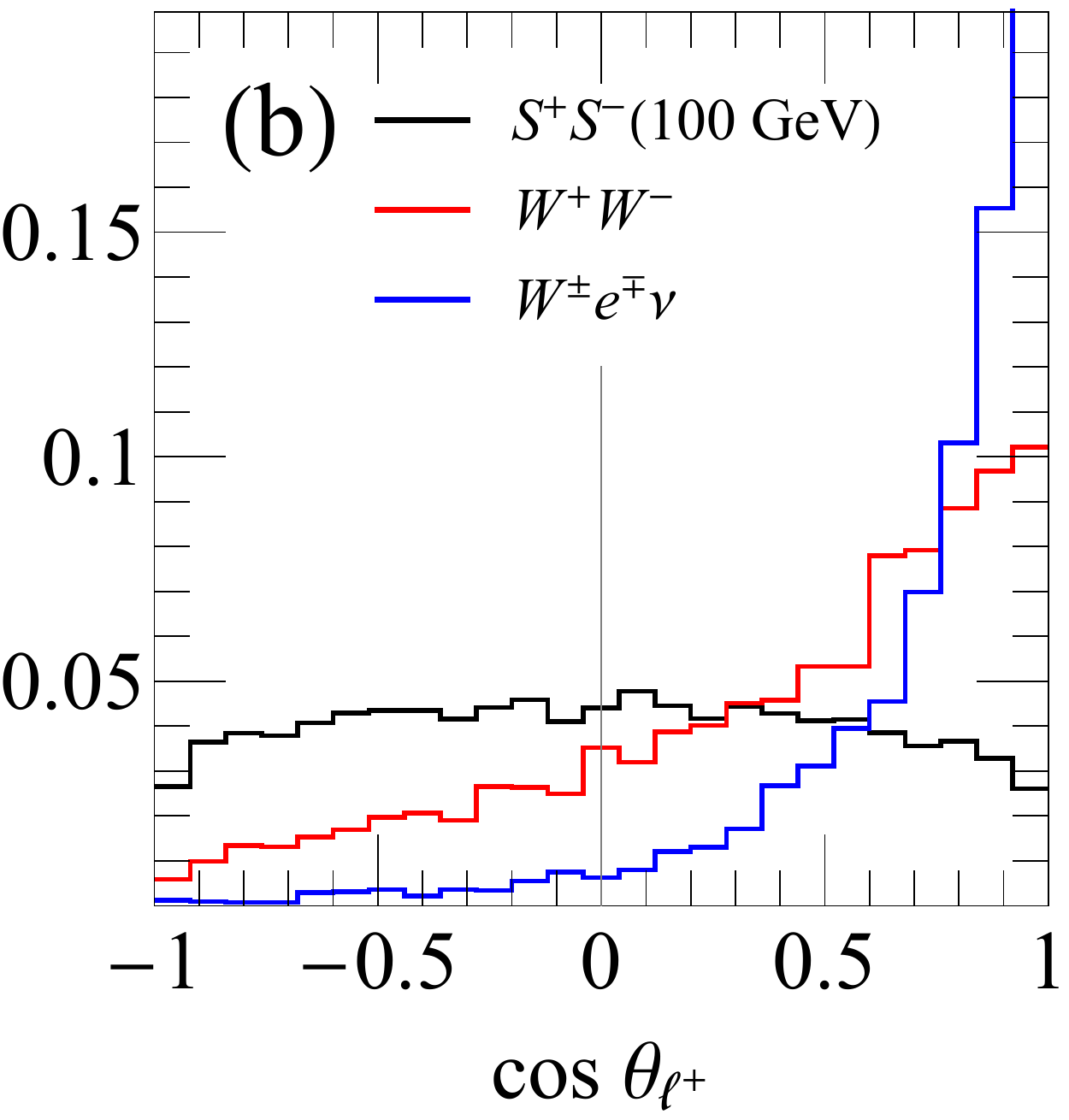}
\caption{The normalized distributions of (a) $M_{T2}$ and (b) $\cos\theta_{\ell^+}$ in the DF dilepton channel after the selection cut. }
\label{fg:lvlv}
\end{figure}

At this stage of the analysis, the background rate is an order of magnitude larger than the signal rate. Moreover, the
dominant background comes from the $W^+ W^-$ process, followed by the $\tau^+\tau^-$ process, $W^{\pm}\ell^{\mp}\nu$ processes, $e^+e^-\ell^+\ell^-$ process, and the $e^+e^-\tau^+\tau^-$ process. The $ZZ$ and $Z\ell^+\ell^-$ processes are negligible. To detect the signal event, further kinematic cuts are needed. In order to study the efficient cuts that can largely suppress the background rates while keeping most of the signal rates, we examine the distributions of the $M_{T2}$  and angular distribution of the charged lepton with respect to the beamline direction. Their normalized distributions are shown in Fig.~\ref{fg:lvlv}.  

The $M_{T2}$ event variable is designed to bound the masses of a
pair of heavy particles that subsequently decay into one or more visible states and missing energy. It is 
a function of the momenta of two visible particles and the missing transverse momentum in an event~\cite{Lester:1999tx}.  Strictly speaking, $M_{T2}(a,b,\met)$ is the minimum of a function
\begin{equation}
\max \left\{ M_T(\vec{p}_T^{~a},\vec{\not{p}}_1),M_T(\vec{p}_T^{~b},\vec{\not{p}}_2)\right\} ,
\end{equation}
such that $\vec{\not{p}}_{1,T}+\vec{\not{p}}_{2,T}=\vec{\not{\!\rm E}}_T$. Here $a$ and $b$ are the two individual (or clustered) visible states from the parent decay, and $\vec{\not{p}}_1, \vec{\not{p}}_2$ are the associated missing transverse momenta.  
The transverse mass $M_{T}$ is defined as 
\begin{equation}
M_{T}(X,\vec{p}_T^{\rm~invis}) = \sqrt{m_X^2 + 2 ( E_T^X E_T^{\rm invis} - \vec{p}_T^X \cdot \vec{p}_{T}^{\rm~invis}) },
\end{equation}
where $X$ denotes the visible particle or cluster. In this study $m_X=0$ for massless neutrinos. The value of $M_{T2}$ variable for each event represents a lower limit of the mass of those heavy particles in the intermediate state, i.e. $m_S$ in our study. Therefore, the $M_{T2}$ distribution exhibits a endpoint around the true mass of intermediate particles. For example, for a 100~GeV charged scalar,  the $M_{T2}$ distribution of the $S^+ S^-$ signal process exhibits a endpoint around $m_S \sim 100~{\rm GeV}$; see the black curve in Fig.~\ref{fg:lvlv}(a). The dominant background $W^+ W^-$ exhibit a endpoint around $m_W\sim 80~{\rm GeV}$; see the red curve.  The $e^+e^-\ell^+\ell^-$ and $\tau^+\tau^-$ processes exhibit much smaller endpoints in the $M_{T2}$ distributions as the two processes do not involve massive gauge bosons in the intermediate state. On the other hand, the $W^\pm e^\mp \nu$ process has a long tail in the $M_{T2}$ distribution owing to the off-shell $W$ boson.

Making use of the difference of $M_{T2}$ distributions, we impose a cut on $M_{T2}$, 
\beq
M_{T2}>20~{\rm GeV},
\eeq
to remove the $\tau^+\tau^-$ and $e^+e^-\ell^+\ell^-$ backgrounds. In Table~\ref{tb:dilepton_DF_CEPC}, we show the number of the signal and background events after the $M_{T2}$ cut. This cut increases the signal-to-background ratio by about a factor of 1.3 while keeping about  one third of the signal rate. The biggest reduction in the background rate comes from the $\tau^+\tau^-$ and $e^+e^-\ell^+\ell^-$ events. 

\begin{table*}
\scriptsize\caption{Cross sections (in the unit of femtobarn) of the signal and backgrounds in the $e^+e^- +\met$ model (left) and $\mu^+\mu^- +\met $ mode (right) at the CEPC. The kinematic cuts listed in each column are applied sequentially.}
\label{tb:SF_ee_mumu_CEPC}

\centering
\begin{minipage}[t]{0.48\textwidth} 
\begin{tabular}{c | c | c | c | c|c|}
\hline
$e^+e^-+\met$ & No cut & Selection & $M_{T2}$ cut & $\cos\theta_{\ell^\pm}$ cut& $\dfrac{\sigma_{S}}{\sqrt{\sigma_B}}\sqrt{\dfrac{\mathcal{L}}{{\rm fb}}} $\\
\hline
(A) & 236.6 & 34.3 & 24.2 & 12.3 & 2.75\\
\hline
(B) & 90.3 &12.8 &10.4	 & 4.6 &1.02 \\
\hline
(C) & 90.3 & 1.5 & 0.6 & 0.3 & 0.07 \\
\hline
\hline
 $W^+W^-$ & 16520 & 143.0 & 103.4 & 17.9 & -\\
 \hline
 $ZZ$ & 1100 & 0.4 & 0.2	& 0.1 & -\\
 \hline
$W^\pm e^\mp \nu$ & 906 & 40.5 & 28.2 & 1.4 & -\\
 \hline
$Z\ell^+\ell^-$ & 527 & 17.9 & 8.8 & 0.6&-\\
 \hline
$e^+e^-\ell^+\ell^-$ (VBS) & 22740 & 287.7 & 0 & 0 &- \\
 \hline
$e^+e^-\tau^+\tau^-$ (VBS)  &5038 & 14.6 & 0 & 0&- \\
 \hline
$\tau^+\tau^-$ & 4321 & 39.5 & 0 & 0 &- \\
 \hline
 \end{tabular}

\end{minipage}
\begin{minipage}[t]{0.5\textwidth} 
\begin{tabular}{|c | c | c | c | c|c}

\hline
$\mu^+\mu^-+\met$ & No cut& Selection & $M_{T2}$ cut & $\cos\theta_{\ell^\pm}$ cut &$\dfrac{\sigma_{S}}{\sqrt{\sigma_B}}\sqrt{\dfrac{\mathcal{L}}{{\rm fb}}} $ \\
\hline
(A) & 236.6 & 46.5 & 32.6 & 17.4 & 3.54 \\
\hline
(B) & 90.3 & 17.5 & 14.1 & 6.6&1.34 \\
\hline
(C) & 90.3 &	1.7 &	 0.7	& 0.4 & 0.08\\
\hline
\hline
 $W^+W^-$ & 16520 & 195.2 & 140.0 & 22.6& -\\
 \hline
 $ZZ$ & 1100 & 0.5 & 0.2	& 0.1&-\\
 \hline
$W^\pm\mu^\mp \nu$ &  51 & 4.9 & 3.7 & 0.7 & -\\
 \hline
$Z\ell^+\ell^-$ & 527 & 5.8 & 3.8 & 0.8 &-\\
 \hline
$e^+e^-\ell^+\ell^-$ (VBS) & 22740 & 356.5 & 0 & 0 &- \\
 \hline
$e^+e^-\tau^+\tau^-$ (VBS)  &5038 & 1.4 & 0 & 0 &-\\
 \hline
$\tau^+\tau^-$ & 4321 & 50.2 & 0 & 0 &- \\
 \hline

\end{tabular}
\end{minipage}

\end{table*}

Another big difference between the signal and the background event signatures is the polar angle distribution of the charged lepton. The polar angle of the charged lepton $\theta_{\ell^\pm}$ is defined as the open angle between the charged lepton $\ell^\pm$ and the motion direction of the incoming positron beams. Fig.~\ref{fg:lvlv}(b) displays the distribution of $\cos\theta_{\ell^+}$ of both the signal and the dominant background processes including $W^+ W^-$, $W^\pm e^\mp \nu$ and $e^+e^-\ell^+\ell^-$. While the background events are populated more often along the $e^+$ beamline, i.e. peaking around $\cos\theta_{\ell^+}\sim 1$, the signal events tend to have a flat $\cos\theta_S$ distribute evenly in the space. The difference can be easily understood as follows. 
First, as explained above, the angular distribution of the charged scalar $S^\pm$ is determined by the Wigner $d$-function $d^{1}_{1,0}(\theta_S)\propto \sin\theta_S$, favoring the central region of the detector. Owing to the scalar feature of $S^\pm$, the decay products of $S^\pm$ distribute isotropically in the space, thus leading to the flat angular distribution. Second, the major backgrounds are produced through the weak interaction which enforces a spin correlation among the initial and final state particles. For example, the $W^+W^-$ pair are produced through two $s$-channel diagrams mediated by $\gamma/Z$ and also a $t$-channel diagram. At a high center of mass energy, the dominant contribution arises from the $t$-channel diagram, which renders the $W^\pm$ bosons favor the forward region due to the factor of $1/(1-\cos\theta_{W^+})$~\cite{Hagiwara:1986vm}. Furthermore, we identify that the $W^+$ and $W^-$ bosons are mainly right-handed and left-handed polarized with respect to the direction of its motion, respectively. As a result, the charged lepton $\ell^+$ from the $W^+$ decay is boosted, resulting in the peak of $\cos\theta_{\ell^+}$ distribution in the forward region. The backgrounds of $W^\pm e^\mp \nu$ and $e^+e^-\ell^+\ell^-$ also prefer the charged lepton $\ell^+$ in the forward region.

To further reduce the major backgrounds from $W^+W^-$ and $W^\pm e^\mp \nu$, we impose cuts on $\cos\theta_{\ell^\pm}$ as follows:
\beq
\cos\theta_{\ell^+} < 0.3,\quad\cos\theta_{\ell^-} > - 0.3,
\eeq
where $\theta_{\ell^\pm}$ is the polar angle of the charged lepton $\ell^\pm$ with respect to the $z$-axis (defined as the direction of the $e^+$ beam). It significantly suppresses all the SM backgrounds; for example, less than $20\%$ of $W^+W^-$ background events pass the $\cos\theta_{\ell^\pm}$ cut, and the $e^+e^-\ell^+\ell^-$ and $e^+e^-\tau^+\tau^-$ backgrounds are negligible. The signal-to-background ratio is increased to 65\% in the model (A) and 23\% in the model (B). See the fifth column in Table~\ref{tb:dilepton_DF_CEPC}. The sixth column lists the significance of discovering the $S^+S^-$ pair in the $e^\pm\mu^\mp+\met$ mode at the CEPC with an integrated luminosity of $1~{\rm fb}^{-1}$. The significance of discovery potential is defined as 
\beq
\mathcal{S}\equiv \frac{N_S}{\sqrt{N_B}},
\eeq
with $N_S$ and $N_B$ being the numbers of the signal and background events, respectively. The significance of other luminosities can be easily obtained from the following luminosity scaling, 
\beq
\mathcal{S}=\frac{\sigma_{S}}{\sqrt{\sigma_{B}}}\sqrt{\frac{\mathcal{L}}{1~{\rm fb}}}.
\eeq 
For the model (A), $m_S=70~{\rm GeV}$ and $\mathcal{B}_e=\mathcal{B}_\mu=0.5$, one can discover the $S^+S^-$ signal at the $4.7\sigma$ level at the CEPC with an integrated luminosity of $1~{\rm fb}^{-1}$. The model (B), $m_S=100~{\rm GeV}$ and $\mathcal{B}_e=\mathcal{B}_\mu=0.5$, can be discovered at the $5\sigma$ level with an integrated luminosity of $\sim 7~{\rm fb}^{-1}$. If the $S^\pm$ decays completely into the $\tau\nu$ mode, then a large luminosity is needed to overcome the suppression of branching ratio of $\tau^\pm\to \ell^\pm \nu$ to reach $5\sigma$ discovery; for example, an integrated luminosity of $4000~{\rm fb}^{-1}$ is needed to reach $5\sigma$ discovery for the model (C). As the CEPC is expected to collect an integrated luminosity of $5~{\rm ab}^{-1}$, it is very promising to observe the singlet charged scalar pairs in the $e^\pm\mu^\mp+\met$ mode at the CEPC as long as the charged scalar predominantly decays into leptons. 

\subsubsection{The SF case: $e^+e^- (\mu^+\mu^-)+\met$}

Next, consider the case of SF leptons.  A big difference from the DF case is the background treatment. In the selection cuts, other than those listed in Eq.~(\ref{eq:selection_cuts}), we require the invariant mass of two leptons to be away from $Z$-pole in the SF channel, i.e., $m_{\ell\ell} \notin [80, 100]~\mathrm{GeV}$. 

In order to get a more realistic simulation, we distinguish the electrons from muons in the analysis. We apply exactly the same set of cuts as previous DF study to obtain the LHC sensitivity. Table~\ref{tb:SF_ee_mumu_CEPC} displays the cross section of the signal and background processes before and after a series of cuts imposed in sequence; the left panel is for the mode of $e^+e^- +\met$ while the right panel for the mode of $\mu^+\mu^- + \met$. More muon events survive and yields better significances in the muon channel in comparison with the electron channel; see the last columns in both the left and right panels. The significance of the SF case is about one half of that of the DF case, mainly owing to the combinatorial factor of $S^\pm$ decay branching ratios.   

For the model (A), $m_S=70~{\rm GeV}$ and $\mathcal{B}_e=\mathcal{B}_\mu=0.5$, one can discover the $S^+S^-$ signal at the $2.8\sigma$ confidence level in the $e^+e^- + \met$ mode and at the $3.5\sigma$ confidence level in the $\mu^+\mu^- +\met$ mode at the CEPC with an integrated luminosity of $1~{\rm fb}^{-1}$. The model (B), $m_S=100~{\rm GeV}$ and $\mathcal{B}_e=\mathcal{B}_\mu=0.5$, can be discovered at the $5\sigma$ level with an integrated luminosity of $\sim 25~{\rm fb}^{-1}$. If the $S^\pm$ decays completely into the $\tau^\pm\nu$ mode, then a large luminosity is needed to overcome the suppression of branching ratio of $\tau^\pm\to \ell^\pm \nu$ to reach $5\sigma$ discovery. The CEPC designed integrated luminosity, $5~{\rm ab}^{-1}$, could yield $5\sigma$ discovery for the model (C). 

\subsubsection{Mass and spin measurement of the scalar $S^\pm$}

It is very promising to observe an excess in the signal of $\ell^\pm\ell^{\prime \mp}+\met$ on top of the SM backgrounds. However, in order to claim the excess is indeed from a charged scalar, it is crucial to determine the scalar mass and to fully reconstruct the kinematics of $S^\pm$ to verify its spin. Below we demonstrate the CEPC is a perfect machine for that job. 

\begin{figure} 
\centering
\includegraphics[scale=0.32]{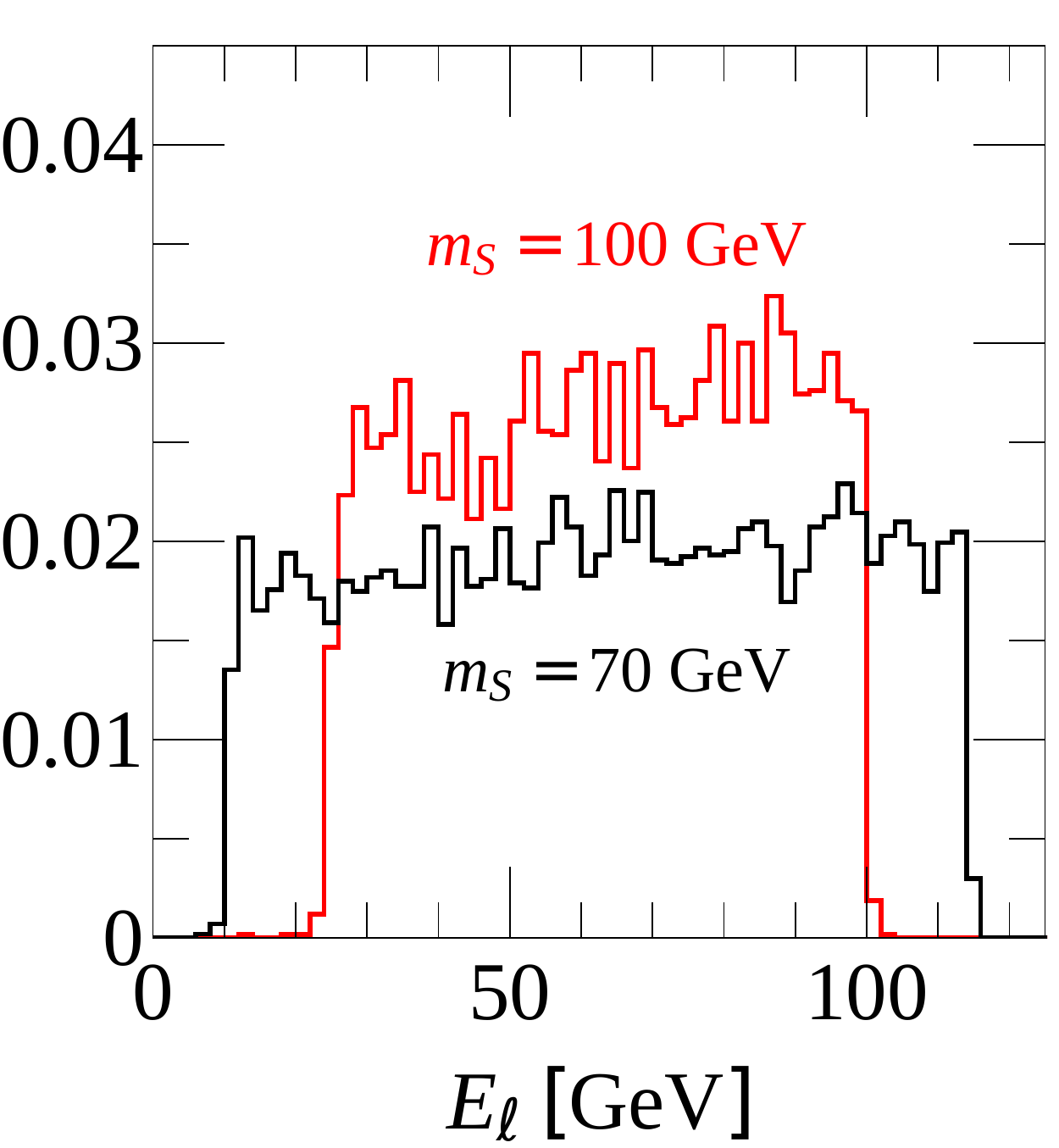}
\caption{The normalized energy distribution of the positively charged lepton in the DF case with $m_S = 70~\mathrm{GeV}$ (black) and $100~\mathrm{GeV}$ (red) after the analysis cuts given in Table~\ref{tb:dilepton_DF_CEPC}.}
\label{fg:precision_dilepton1}
\end{figure}

To determine $m_S$, one can either study the $M_{T2}$ distribution depicted in Fig.~\ref{fg:lvlv}(a), or examine the energy distributions of charged leptons in the final state~\cite{Cao:2007pv}. Figure~\ref{fg:precision_dilepton1} displays the normalized distributions of the energy of the positively charged leptons $\ell^+$ in the DF case with $m_S = 70~\mathrm{GeV}$ (black) and $100~\mathrm{GeV}$ (red) after the analysis cuts listed in Table~\ref{tb:dilepton_DF_CEPC}. Note that choosing the negatively charged leptons makes no difference. Two end points, one large $E_{\ell}^+$ and one small $E_{\ell}^-$, are then observed, and either of them can be used to extract the value of $m_S$. Specifically, the two end points arise from the two extreme scenarios that the charged lepton $\ell^+$ travels in the same or opposite direction of the scalar $S^+$, yielding
\begin{eqnarray}
E_{\ell}^{\pm} = \frac{\sqrt{s}}{4} \pm \frac{1}{2} \sqrt{\frac{s}{4} - m_S^2}~,
\end{eqnarray}
from which we obtain $m_S$ as following,
\begin{eqnarray}
m_S = \sqrt{2 E_\ell^\pm ( \sqrt{s} - 2 E_\ell^\pm)}~.
\end{eqnarray}
Here we ignore the masses of leptons in the final state.

After pinning down $m_S$, one can employ it to reconstruct the full kinematics of each event and further determine the spin of the scalar $S^\pm$ through the angular distributions of the reconstructed charged scalar $S^\pm$ and the charged leptons. Following Refs.~\cite{Hagiwara:1986vm,Cao:2007pv}, we denote the 3-momenta of the positively and negatively charged leptons as $\vec{p}_{\ell^+}$ and $\vec{p}_{\ell^-}$, respectively, the 3-momentum of the final state neutrino $\vec{p}_\nu$ can then be decomposed as
\begin{eqnarray}
\vec{p}_\nu = A \vec{p}_{\ell^+} + B \vec{p}_{\ell^-} + C \vec{p}_{\ell^+} \times \vec{p}_{\ell^-}~.
\label{eq:pnu}
\end{eqnarray}
The coefficients of $A$ and $B$ are determined by
\begin{eqnarray}
\begin{pmatrix}
A \\ B
\end{pmatrix} = \frac{1}{L} \begin{pmatrix}
|\vec{p}_{\ell^-}|^2 & -\vec{p}_{\ell^+_{}} \cdot \vec{p}_{\ell^-_{}} \\
-\vec{p}_{\ell^+_{}} \cdot \vec{p}_{\ell^-_{}} & |\vec{p}_{\ell^+}|^2
\end{pmatrix} \begin{pmatrix}
M \\ N
\end{pmatrix}~,
\end{eqnarray}
with $L$, $M$ and $N$ given by
\begin{eqnarray}
L & \equiv &  |\vec{p}_{\ell^+}|^2 |\vec{p}_{\ell^-}|^2 - (\vec{p}_{\ell^+_{}} \cdot \vec{p}_{\ell^-_{}})^2_{} , \nonumber \\
M & \equiv & \frac{1}{2}  \left[ E_{S^+}^2 - m_S^2 - E_{\ell^+}^2  -E_\nu^2 \right], \nonumber \\
N & \equiv & \frac{1}{2}  \left[ E_{\bar{\nu}}^2 - E_{S^-}^2 + m_S^2  - E_{\ell^-}^2 - 2 \vec{p}_{\ell^+_{}} \cdot \vec{p}_{\ell^-_{}}  \right],
\end{eqnarray}
where $E_{S^\pm}$ denotes the energy of the scalar $S^\pm$, and $E_{\nu} $ ($E_{\bar{\nu}}$) is the energy of the final state neutrino (anti-neutrino), respectively. The remaining coefficient $C$ can be found as
\begin{eqnarray}
C^2 &=& \frac{1}{|\vec{p}_{\ell^+} \times \vec{p}_{\ell^-}|^2} \left[ E_\nu^2 - A^2 |\vec{p}_{\ell^+}|^2 \right. \nonumber \\
& & \left. - B^2  |\vec{p}_{\ell^-}|^2 - 2 A B~\vec{p}_{\ell^+_{}} \cdot \vec{p}_{\ell^-_{}} \right]~.
\end{eqnarray}
The sign of $C$ is ambiguious, and here we always take $C > 0$. It is straightforward to fully reconstruct the kinematics of the charged scalar $S^\pm$ once solving $\vec{p}_\nu$ from Eq.~(\ref{eq:pnu}).

\begin{figure}
\centering
\includegraphics[scale=0.32]{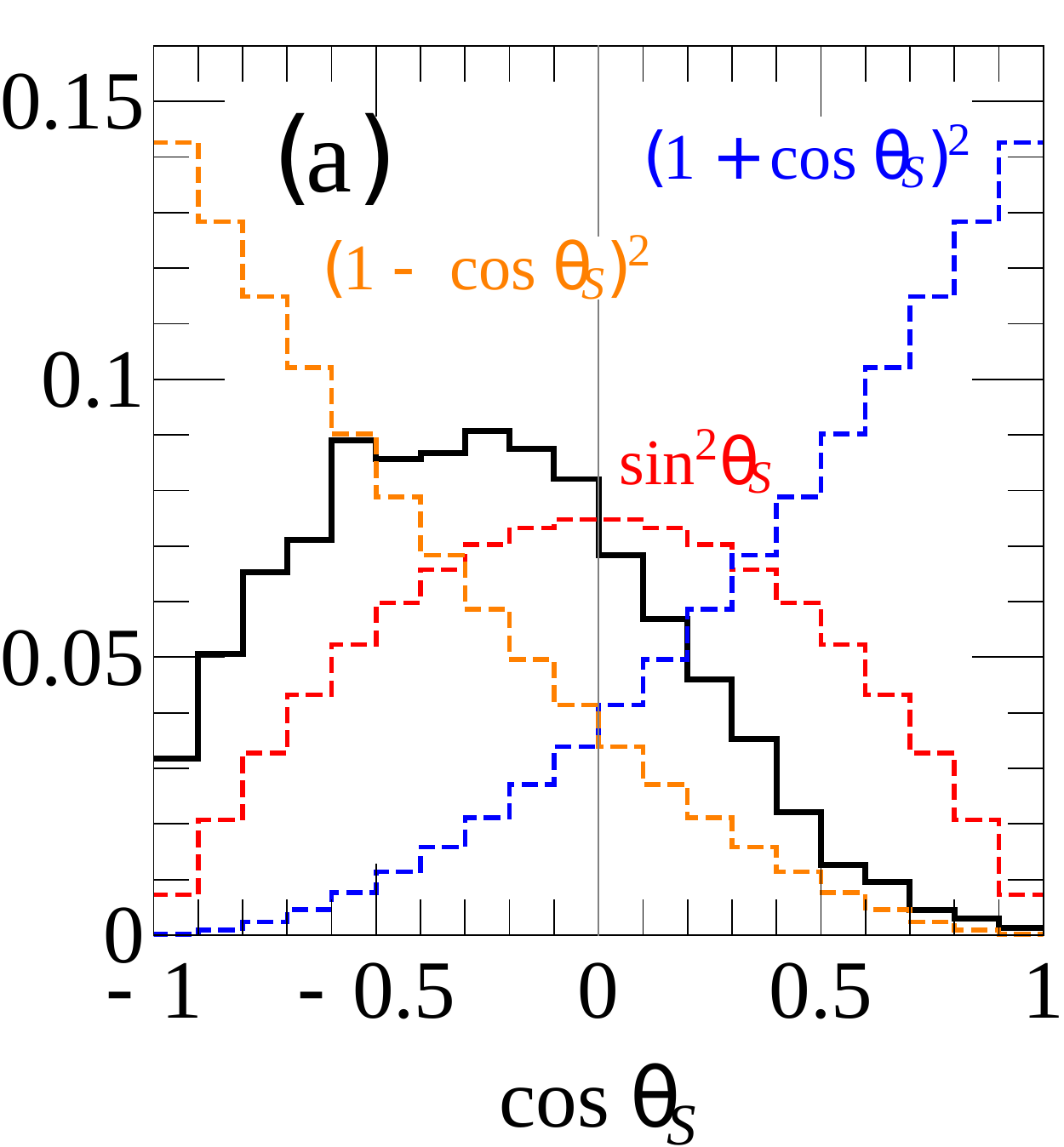}
\includegraphics[scale=0.32]{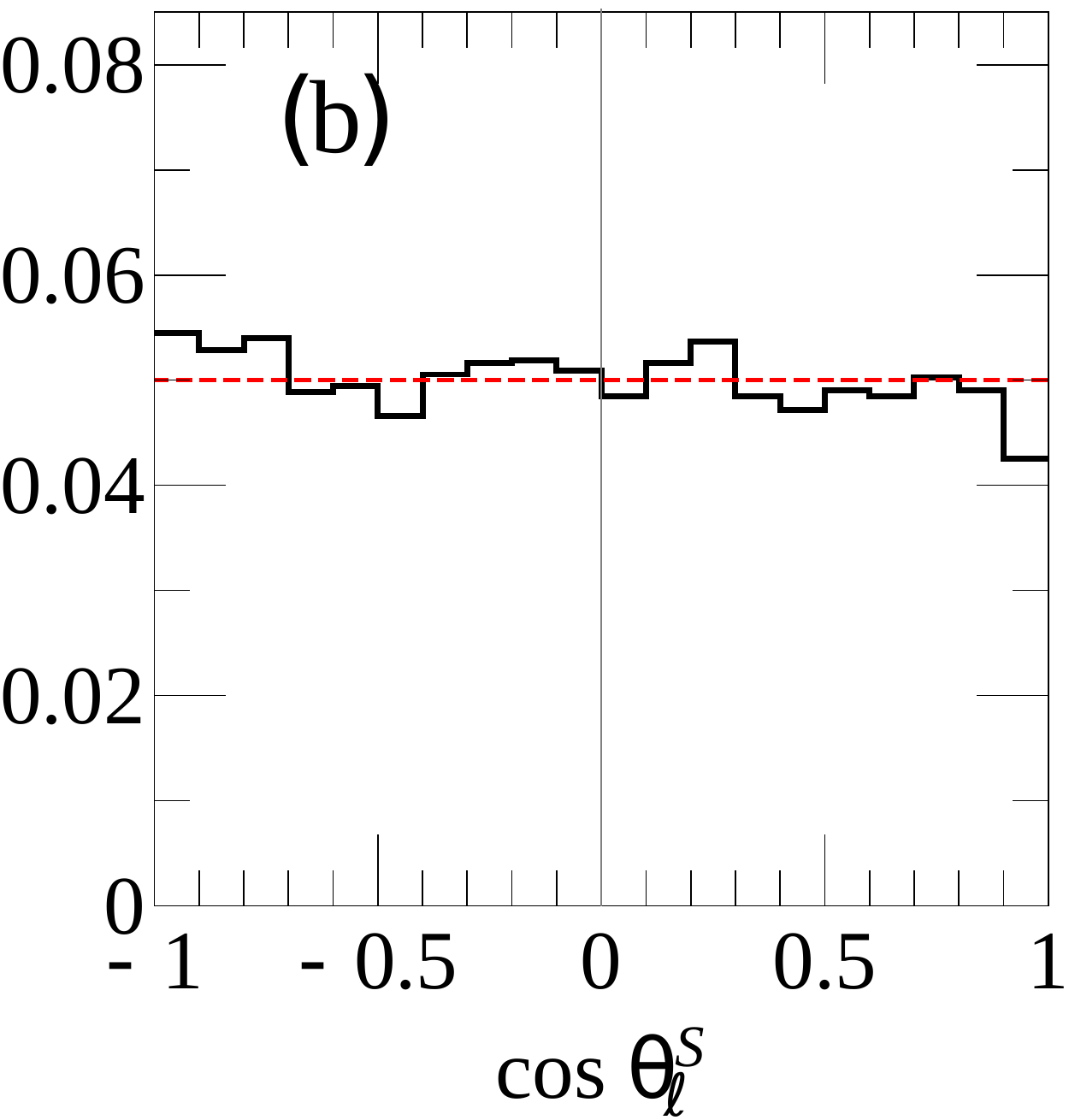}
\caption{(a) Normalized distribution of $\cos\theta_S$; (b) normalized distribution of $\cos\theta_\ell^S$, where $\theta_\ell^S$ is defined as the polar angle of the positively charged lepton in the rest frame of its mother particle $S^+$ with the $z$-axis chosen as the moving direction of $S^+$ in the lab frame. Red dashed line is the expected distribution from a scalar decay.}
\label{fg:precision_dilepton}
\end{figure}

Figure~\ref{fg:precision_dilepton}(a) shows the normalized distribution of $\cos\theta_S$ in the DF case with $m_S = 70~\mathrm{GeV}$ (black), where $\theta_S$ is the polar angle of the reconstructed scalar $S^+$ with respect to the beamline direction. For comparison, we also plot three benchmark distributions of $\sin^2\theta_S$ (red dashed), $(1+\cos\theta_S)^2$ (blue dashed) and $(1-\cos\theta_S)^2$ (orange dashed), which correspond to the square of the Wigner $d$-functions $d_{1,0}^1(\theta_S)$, $d_{1,1}^1(\theta_S)$ and $d_{1,-1}^1(\theta_S)$, respectively. The signal process mostly resembles the distribution of $\sin^2\theta_S$, agreeing with what one would expect from a pair production of scalars. The asymmetric distortion of the signal process comparing with that of $\sin^2\theta_S$ is mainly due to the cut $\cos\theta_{\ell^+} < 0.3$, which tends to suppress events with large values of $\cos\theta_S$.  

Finally, in Fig.~\ref{fg:precision_dilepton}(b) we plot the normalized distribution of $\cos\theta_\ell^S$ (black), where $\theta_\ell^S$ is the polar angle of the positively charged lepton in the rest frame of $S^+$ ($z$-axis defined as the moving direction of $S^+$ in the lab frame). Again, the nearly flat shape confirms the scalar nature of $S^\pm$. 

\subsection{The $\tau^+\tau^-+\met$ mode} 

The analysis of the tau leptons is slightly more complicated than that of the electron and muon leptons as the tau lepton would decay into leptons or hadrons inside the detector. We focus on the so-called ``one-prong" tau decays which are selected by choosing tau decay cones containing only one well-reconstructed charged track, consistent with coming from the origin. In the $\tau^+\tau^-\nu\bar{\nu}$ channel we only consider the hadronic decay modes of $\tau$ leptons, and for simplicity, we employ the reconstructed $\tau$-jets from detector simulation. In the selection cuts we demand exactly two tagged $\tau$-jets ($\tau_h$) satisfying 
\begin{eqnarray} \label{eq:tau_jet_region}
p_T^{\tau_h} > 5~\mathrm{GeV}, \quad |\eta^{\tau_h} |  < 3,
\end{eqnarray}
and veto any electrons, muons or other QCD jets in the central region of $|p_T| > 5~\mathrm{GeV}$ and $|\eta| < 3$. The missing energy is required to be $$\met > 5~\mathrm{GeV}.$$ The cross sections of the signal and background processes after the selection cuts are given in Table~\ref{tb:tata_CEPC_tau_0p6}. At this stage the dominant background comes from the $\tau^+\tau^-$ process. To suppress it, we again apply a $M_{T2}$ cut, i.e., $M_{T2} > 20~\mathrm{GeV}$, as in the previous dilepton analysis.

\begin{table}[b]
\scriptsize
\centering
\caption{Cross sections (in the unit of femtobarn) of the signal and backgrounds in the $\tau^+\tau^-\nu\bar{\nu}$ mode at the CEPC. The kinematic cuts listed in each column are applied sequentially.}
\label{tb:tata_CEPC_tau_0p6}
\begin{tabular}{c | c | c | c | c  }
\hline
$\tau^+\nu\tau^-\bar{\nu}$ & No cut & Selection & $M_{T2}$ cut & $\left |\cos\theta_{\tau_h} \right |$ cut \\
\hline
$m_S = 100~\mathrm{GeV}$ &  \multirow{2}{*}{90.3}	&	 \multirow{2}{*}{11.1}	&	 \multirow{2}{*}{7.3}	&	 \multirow{2}{*}{4.9}\\
$\mathcal{B}_\tau  = 1$ & & & & \\
\hline
\hline
$W^+W^-$ & 16520	&	23.7 & 14.0 & 5.8 \\
\hline
$ZZ$ & 1100	&	2.0 & 1.1 & 0.6 \\
\hline
$W^\pm\tau^\mp\nu$ & 51	&	0.7	&	0.4	&	0.2\\
\hline
$Z\tau^+\tau^-$ & 55	&	0.9	&	0.5	&	0.2\\
\hline
$e^+e^-\tau^+\tau^-$ (VBS) & 5038	&	13.3	&	0	&	0 \\
\hline
$\tau^+\tau^-$ & 4321	&	402.0	&	0.2	&	0.2 \\
\hline
$\gamma \gamma \rightarrow \tau^+\tau^-$ & \multirow{2}{*}{59325}	&	\multirow{2}{*}{37.3}	&	\multirow{2}{*}{0}	&	\multirow{2}{*}{0}	 \\
Beamstrahlung & & & &  \\
\hline
\end{tabular}
\end{table}

\begin{figure}
\centering
\includegraphics[scale=0.32]{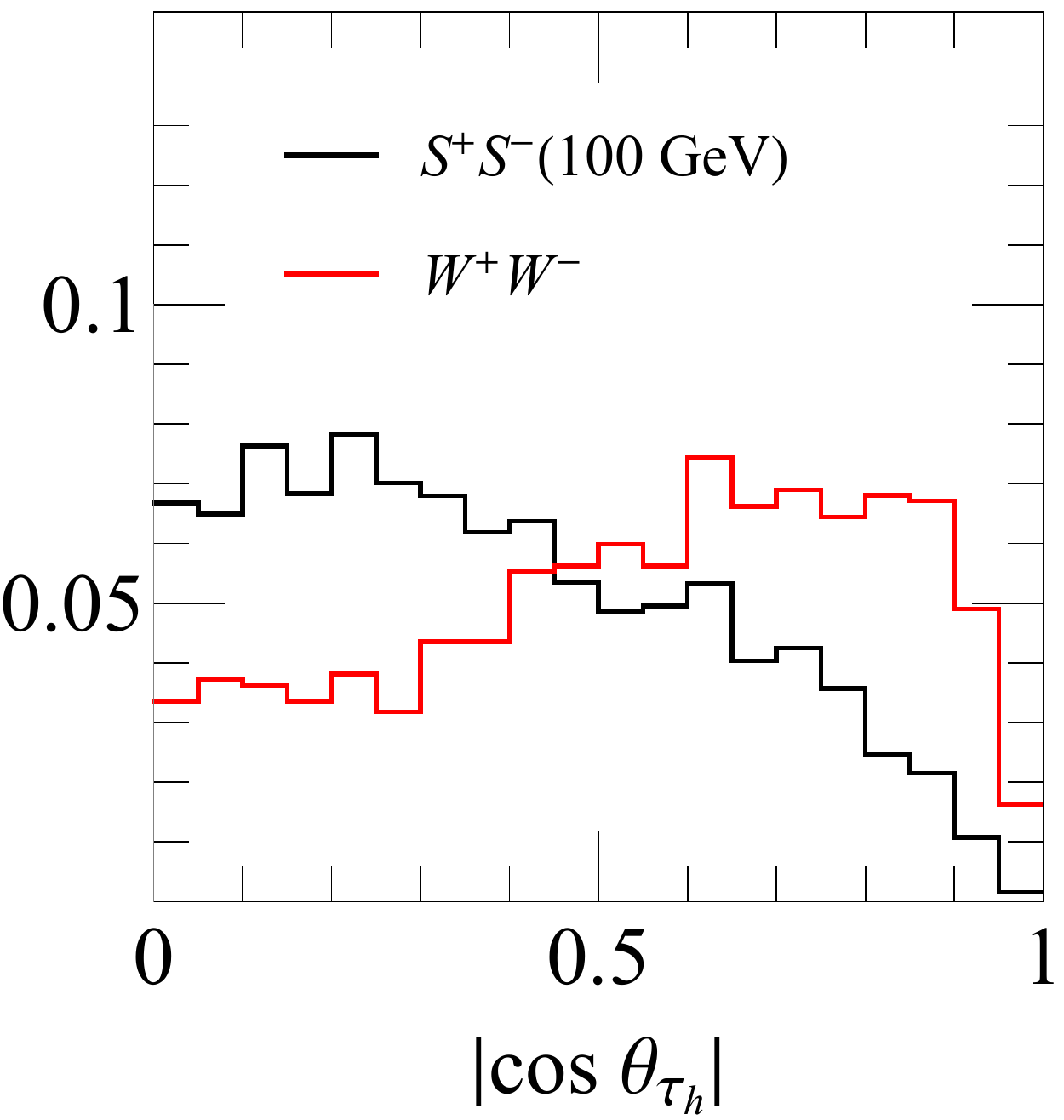}
\caption{Distribution of $|\cos\theta_{\tau_h}|$ in the $\tau^+\tau^-+\met$ channel after the $M_{T2}$ cut.}
\label{fg:tvtv}
\end{figure}

After the $M_{T2}$ cut the $W^+W^-$ process turns out to be the dominant background (see Table~\ref{tb:tata_CEPC_tau_0p6}). Employing the fact that the $\tau$-jets from the decay of $W^{\pm}$ bosons tend to favor the forward region, we further impose the $|\cos\theta_{\tau_h}|$ cut, where $\theta_{\tau_h}$ is the polar angle of the leading $\tau$-jet. In Fig.~\ref{fg:tvtv} we show the distributions of $|\cos\theta_{\tau_h}|$ for the signal process with $m_S = 100~\mathrm{GeV}$ (black) and the $W^+W^-$ background (red). Thus, a cut of 
\begin{eqnarray}
|\cos\theta_{\tau_h}| < 0.5,
\end{eqnarray}
can be used to suppress the $W^+W^-$ background. The last column of Table~\ref{tb:tata_CEPC_tau_0p6} lists the cross sections of the signal and background processes after all the analysis cuts. Hence, for the case of $m_S = 100~\mathrm{GeV}$ with $\mathcal{B}_\tau = 1$ one can reach $5\sigma$ discovery with an integrated luminosity of $7.5~\mathrm{fb}^{-1}$.

\begin{figure}
\centering
\includegraphics[scale=0.32]{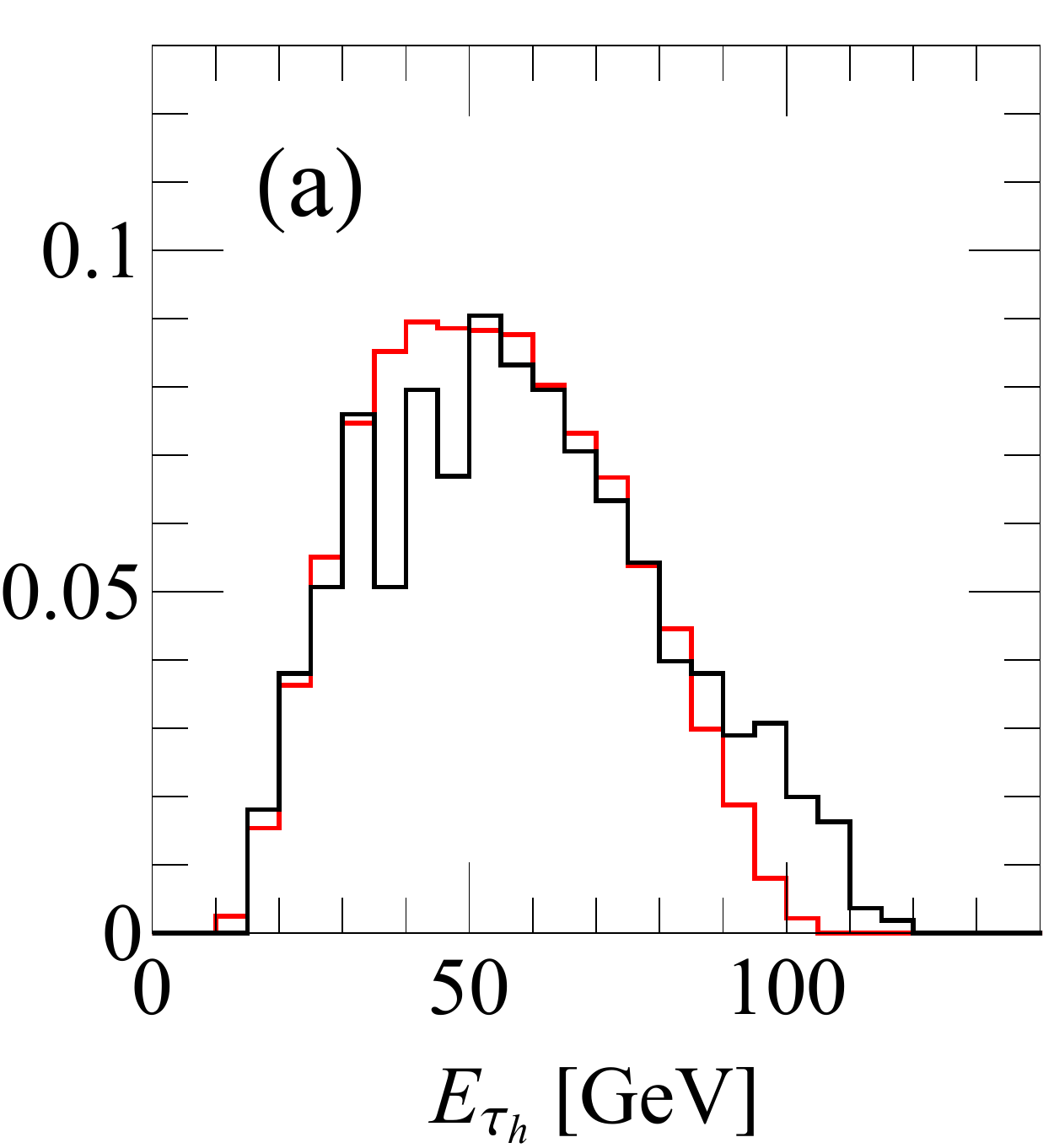}
\includegraphics[scale=0.32]{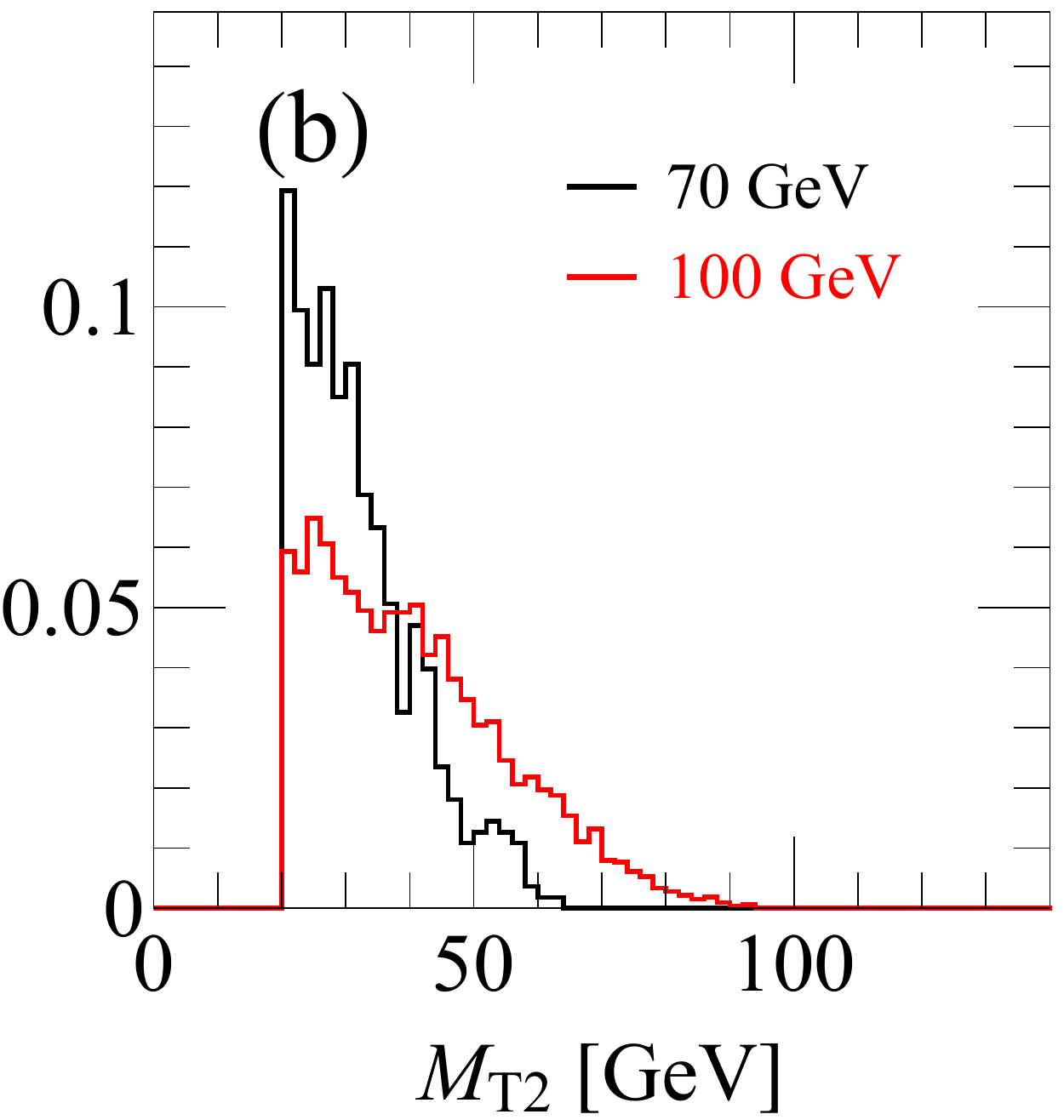}
\caption{Normalized distributions of (a) $E_{\tau_h}$ and (b) $M_{T2}$ for the signal processes with $m_S = 70~\mathrm{GeV}$ (black) and $m_S = 100~\mathrm{GeV}$ (red) in the $\tau^+\nu\tau^-\bar{\nu}$ mode at the CEPC.}
\label{fg:tvtv_precision}
\end{figure}

The precision measurement in the $\tau^+\tau^-\nu\bar{\nu}$ mode can also be carried out in the post-discovery era. In Fig.~\ref{fg:tvtv_precision} we plot the distributions of $E_{\tau_h}$ (left) and $M_{T2}$ (right) for the signal processes with $m_S = 70~\mathrm{GeV}$ (black) and $m_S = 100~\mathrm{GeV}$ (red) after the cuts shown in Table~\ref{tb:tata_CEPC_tau_0p6}. Here, $E_{\tau_h}$ is the energy of the leading $\tau$-jet. Comparing with the previous dilepton case, since the decay of $\tau$-leptons produces neutrinos, the end points in the distributions of both $E_{\tau_h}$  and $M_{T2}$ are less prominent, posing a challenge in the determination of $m_S$. Moreover, because of the presence of neutrinos from the $\tau$ lepton decay, there exist some discrepancies in the kinematics between the reconstructed $\tau$-jets and the true $\tau$-leptons. As a result, it is hard to employ the method introduced in the previous dilepton analysis to reconstruct the kinematics of charged scalar $S^\pm$ using the momenta of the $\tau$-jets. More advanced analysis techniques are needed in the $\tau^+\tau^-\nu\bar{\nu}$ channel.

\subsection{The $\tau^\pm jj +\met$ mode} 

Similar to the previous $\tau^+\tau^-\nu\bar{\nu}$ mode, we consider the hadronic decay modes of $\tau$-leptons in the $\tau^\pm\nu jj$ mode. In the selection cuts, we require one tagged $\tau$-jet in the same central region as Eq.~(\ref{eq:tau_jet_region}), and two other QCD jets, which are not tagged as $\tau$-jets, satisfying $p_T > 5~\mathrm{GeV}$ and $|\eta|  < 3$. Moreover, no electrons and muons are allowed if they satisfy $p_T > 5~\mathrm{GeV}$ and $|\eta|  < 3$, and the missing energy is required to be $\met > 5~\mathrm{GeV}$. The cross sections of signal and background processes after the selection cuts are presented in the third column of Table~\ref{tb:tajj_CEPC_tau_0p6}, which shows that the dominant background comes from the $W^+W^-$ process.

\begin{table}
\scriptsize
\centering
\caption{Cross sections (in the unit of femtobarn) of the signal and backgrounds in the $\tau^\pm\nu j j$ mode at the CEPC. The kinematic cuts listed in each column are applied sequentially.}
\label{tb:tajj_CEPC_tau_0p6}
\begin{tabular}{c | c | c | c | c  }
\hline
$\tau^\pm\nu j j$ & No cut & Selection & $ |\cos\theta_S| $ cut & $\Phi_{jj}$ cut \\
\hline
$m_S = 100~\mathrm{GeV}$ &  \multirow{2}{*}{90.3}	&	\multirow{2}{*}{10.8}	&	\multirow{2}{*}{8.6	}	&	\multirow{2}{*}{8.0}	 \\
$\mathcal{B}_\tau  = \mathcal{B}_j = 0.5$ & & & & \\
\hline
\hline
$W^+W^-$ & 16520	&	548.2	&	166.5	&	38.9 \\
\hline
$ZZ$ & 1100	&	4.5	&	2.1 & 1.1 \\
\hline
$Zh$ & 212	&	1.8 & 1.2 & 1.0 \\
\hline
$W^\pm\tau^\mp\nu$ & 51	&	8.2 	&	3.5 & 2.3 \\
\hline
$W^\pm q\bar q^\prime$ & 307	&	6.6 & 2.9 & 2.5 \\
\hline
$jjjj$-QCD & 15280	&	14.2	&	8.7 & 4.6 \\
\hline
\end{tabular}
\end{table}

\begin{figure}
\centering
\includegraphics[scale=0.32]{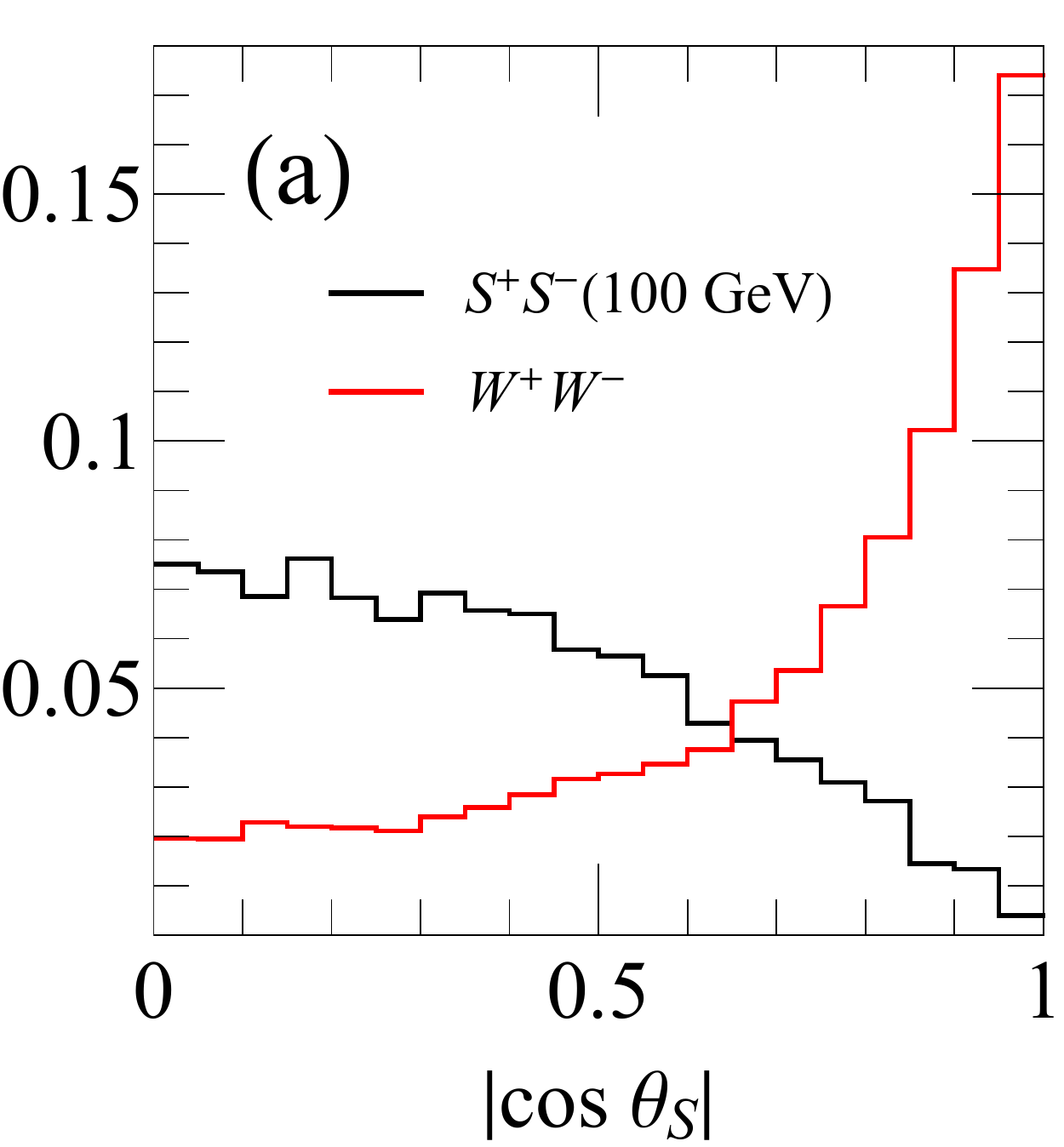}\quad
\includegraphics[scale=0.3]{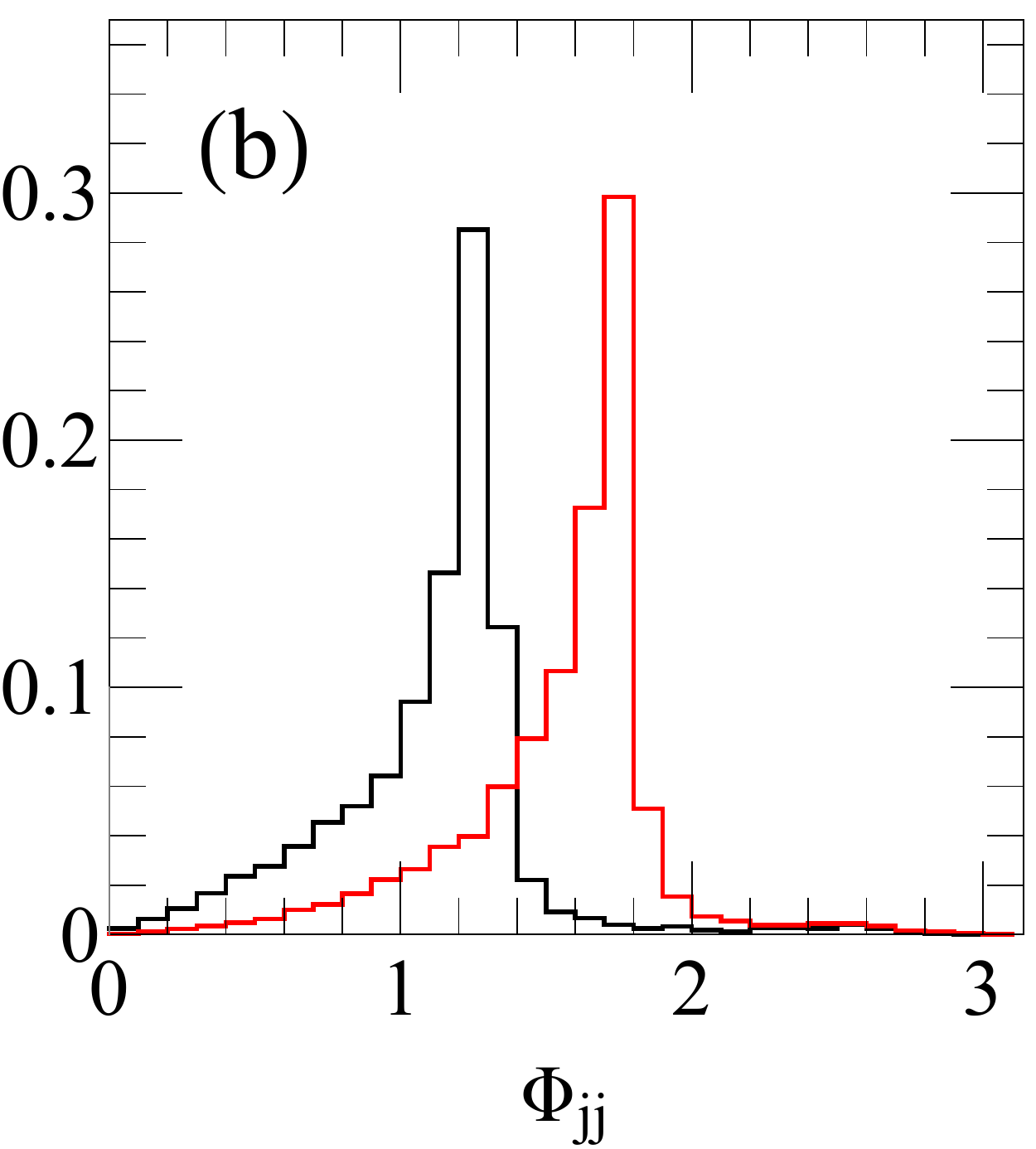}
\caption{Distributions of (a) $|\cos\theta_S| $ and (b) $\Phi_{jj}$ for the signal process with $m_S = 100~\mathrm{GeV}$ (black) and the $W^+W^-$ background (red) in the $\tau^\pm\nu jj$ mode at the CEPC.}
\label{fg:tvjj}
\end{figure}

To suppress the $W^+W^-$ background, we make use of the fact that the $W^\pm$ bosons are produced mostly in the forward region while the scalars $S^\pm$ favor the central region (the charged scalars exhibit a $p$-wave angular momentum). Thanks to the two non-$\tau$-jets in final state, the kinematics of the intermediate $W^{\pm}$ bosons or the scalars $S^\pm$ can be fully reconstructed. We define the variable $\theta_S$ as the polar angle of the intermediate particle in the lab frame, in Fig.~\ref{fg:tvjj}(a) we plot the distribution of $|\cos\theta_S|$ for both the signal process with $m_S = 100~\mathrm{GeV}$ (black) and the $W^+W^-$ background (red). We then observe that the following cut on $\cos\theta_S$, 
\begin{eqnarray}
|\cos\theta_S| < 0.6~,
\end{eqnarray}
is efficient to reduce the $W^+W^-$ background. The cross sections of the signal and background processes after applying the $|\cos\theta_S|$ cut are given in the fourth column of Table~\ref{tb:tajj_CEPC_tau_0p6}.

Finally, to improve the sensitivity to the high mass region of the scalar $S^\pm$, we consider a cut on the kinematic variable $\Phi_{jj}$, defined as the supplement of the open angle between the two non-$\tau$-jets $j_1$ and $j_2$ in the final state, i.e.
\beq
\Phi_{jj}=\pi-\arccos\frac{\vec{p}_{j_1}\cdot\vec{p}_{j_2}}{|\vec{p}_{j_1}||\vec{p}_{j_2}|}.
\eeq
When the mass of the scalar $S^\pm$ increases, the produced scalars $S^\pm$ tends to stay at rest, so that the two jets from the $S^\pm$ decay are likely to be back-to-back in the lab frame, resulting in a smaller value of $\Phi_{jj}$. In Fig.~\ref{fg:tvjj}(b) we plot the distributions of $\Phi_{jj}$ for both the signal process with $m_S = 100~\mathrm{GeV}$ (black) and the $W^+W^-$ background (red). Thus, applying the cut of 
\begin{eqnarray}
\Phi_{jj} < 1.4~,
\end{eqnarray}
can siginificantly reduce the $W^+W^-$ background. Since the $\Phi_{jj}$ cut is only effective for $m_S \gtrsim 100~\mathrm{GeV}$, we consider two scenarios, i.e., with and without the $\Phi_{jj}$ cut, when scanning over different values of $m_S$, and seek the best sensitivity by combining the two scenarios. In addition, other cuts such as the transverse mass of the $\tau$-jet and $\met$ can also be used to suppress the background~\cite{Cao:2003tr}.  

\begin{figure}
\centering
\includegraphics[scale=0.32]{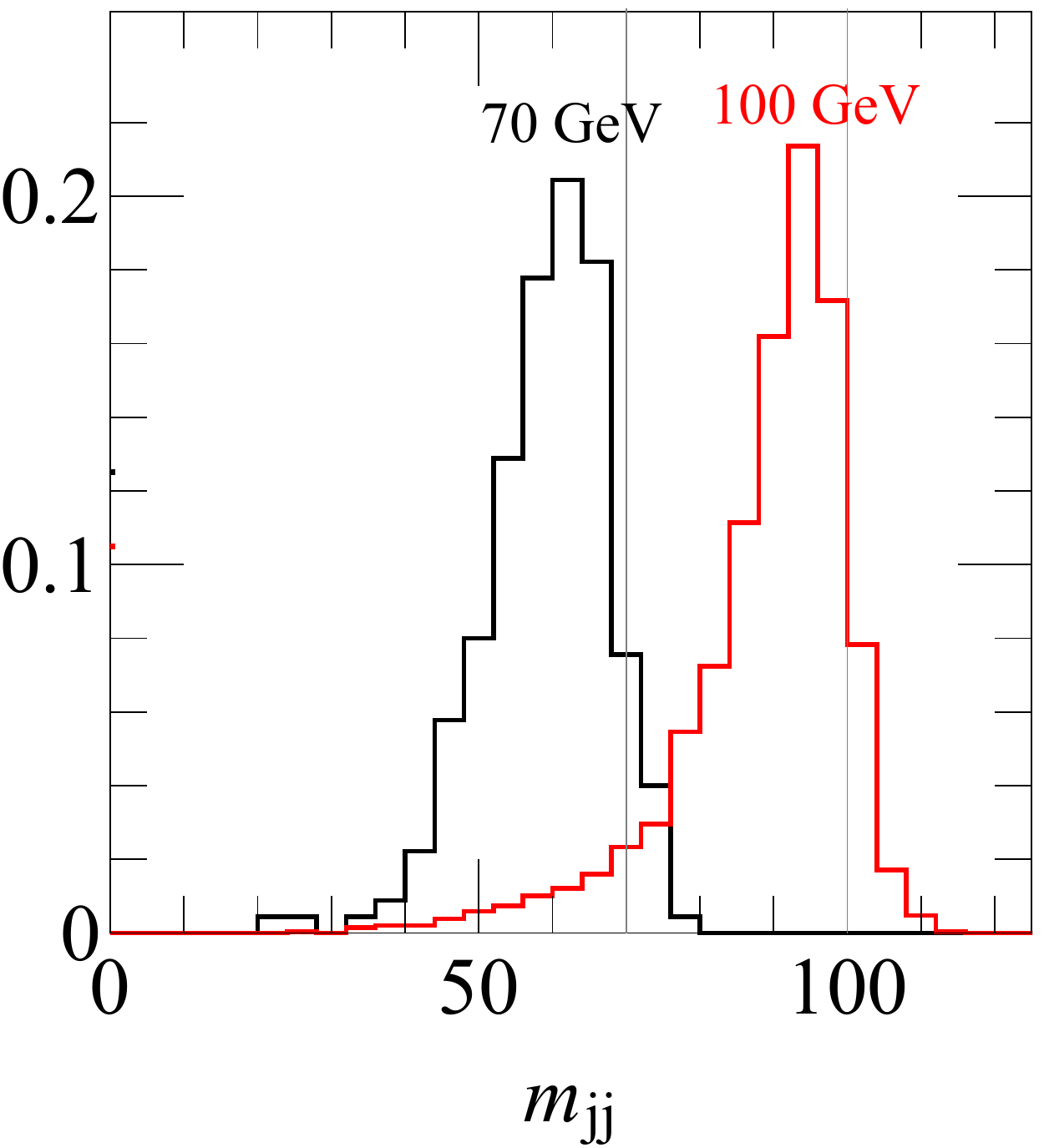}
\caption{Normalized distributions of $m_{jj}$ in the $\tau^\pm\nu jj$ mode for the signal processes with $m_S = 70~\mathrm{GeV}$ (black) and $m_S = 100~\mathrm{GeV}$ (red) after all the cuts given in Table~\ref{tb:tajj_CEPC_tau_0p6}.} 
\label{fg:tajj_precision}
\end{figure}

Regarding the determination of the mass and spin of the scalar $S^{\pm}$ in the $\tau^\pm\nu jj$ mode, we first obtain $m_S$ from the invariant mass $m_{jj}$ of the two non-$\tau$-jets in final state. Figure~\ref{fg:tajj_precision} shows the normalized distributions of $m_{jj}$ in the $\tau^\pm\nu jj$ mode for the signal processes with $m_S = 70~\mathrm{GeV}$ (black) and $m_S = 100~\mathrm{GeV}$ (red) after applying the analysis cuts given in Table~\ref{tb:tajj_CEPC_tau_0p6}. Note that the reconstructed value of $m_S$ is slightly smaller than the true value. It is mainly due to additional QCD radiations in the hadronic decay of the scalar $S^\pm$. 

After reconstructing the kinematics of the scalar $S^\pm$, we can determine the spin of $S^\pm$ in a similar way as that in the previous dilepton analysis. For example, the angular distribtion of the reconstructed scalar $S^\pm$, e.g., the distribution of $|\cos\theta_S|$ shown in Fig.~\ref{fg:tvjj}(a), can reveal the scalar nature of $S^\pm$. One can also plot the angular distribution of the $\tau$-jet in the rest frame of the scalar $S^\pm$ to further confirm the scalar nature of $S^\pm$. 

\subsection{The $jjjj$ mode} 

\begin{figure}
\centering
\includegraphics[scale=0.32]{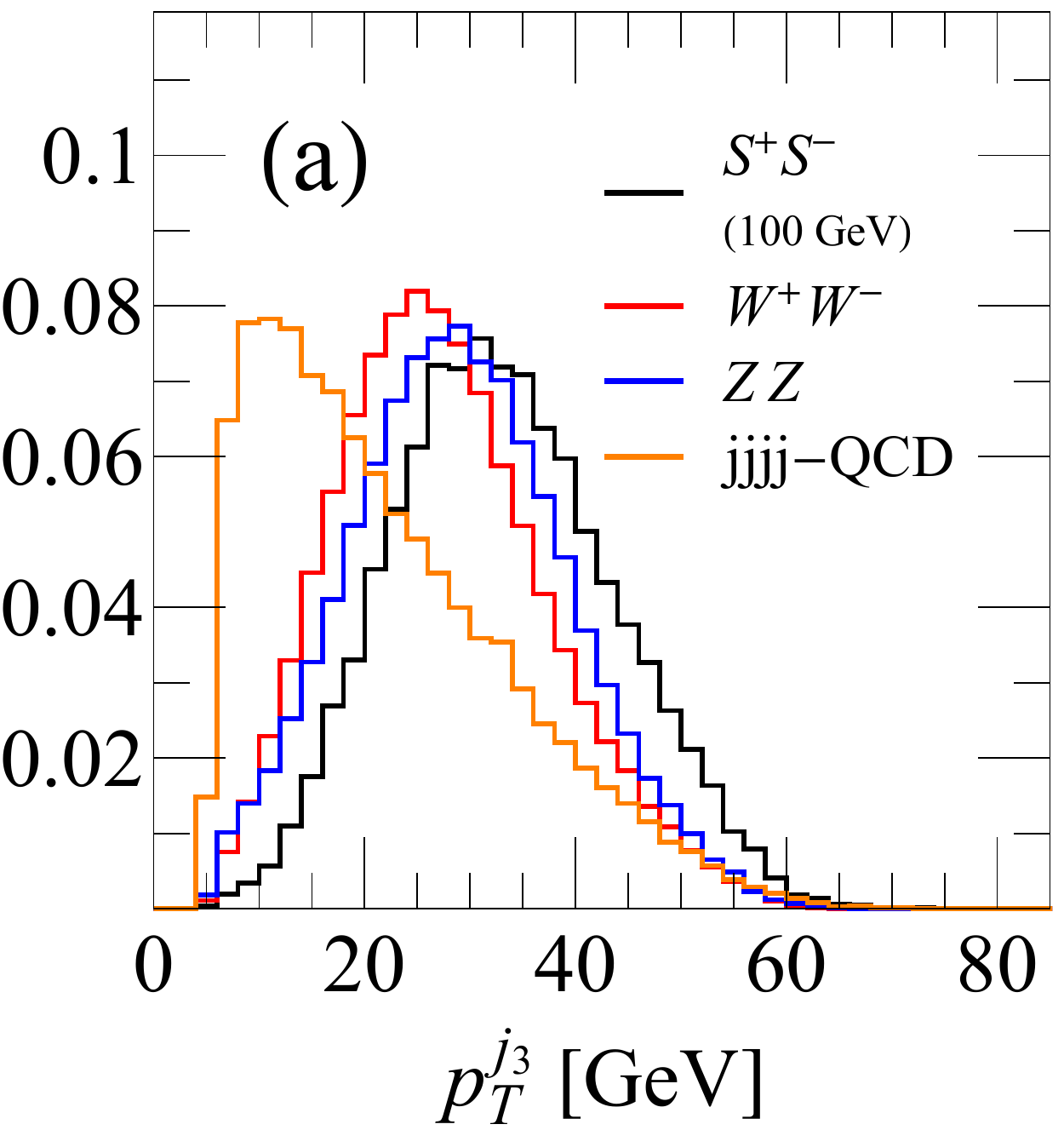}
\includegraphics[scale=0.31]{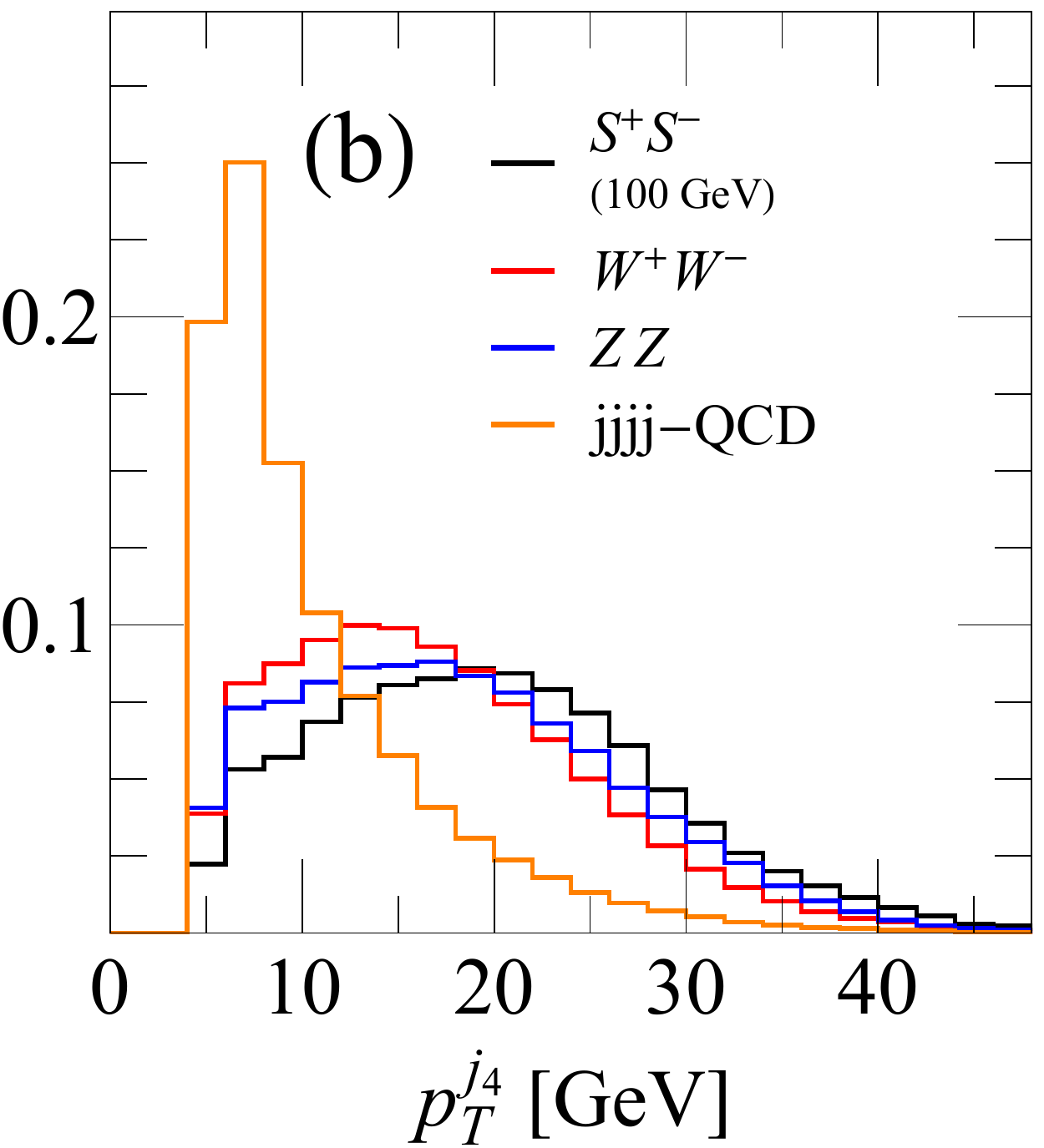}\\
\includegraphics[scale=0.32]{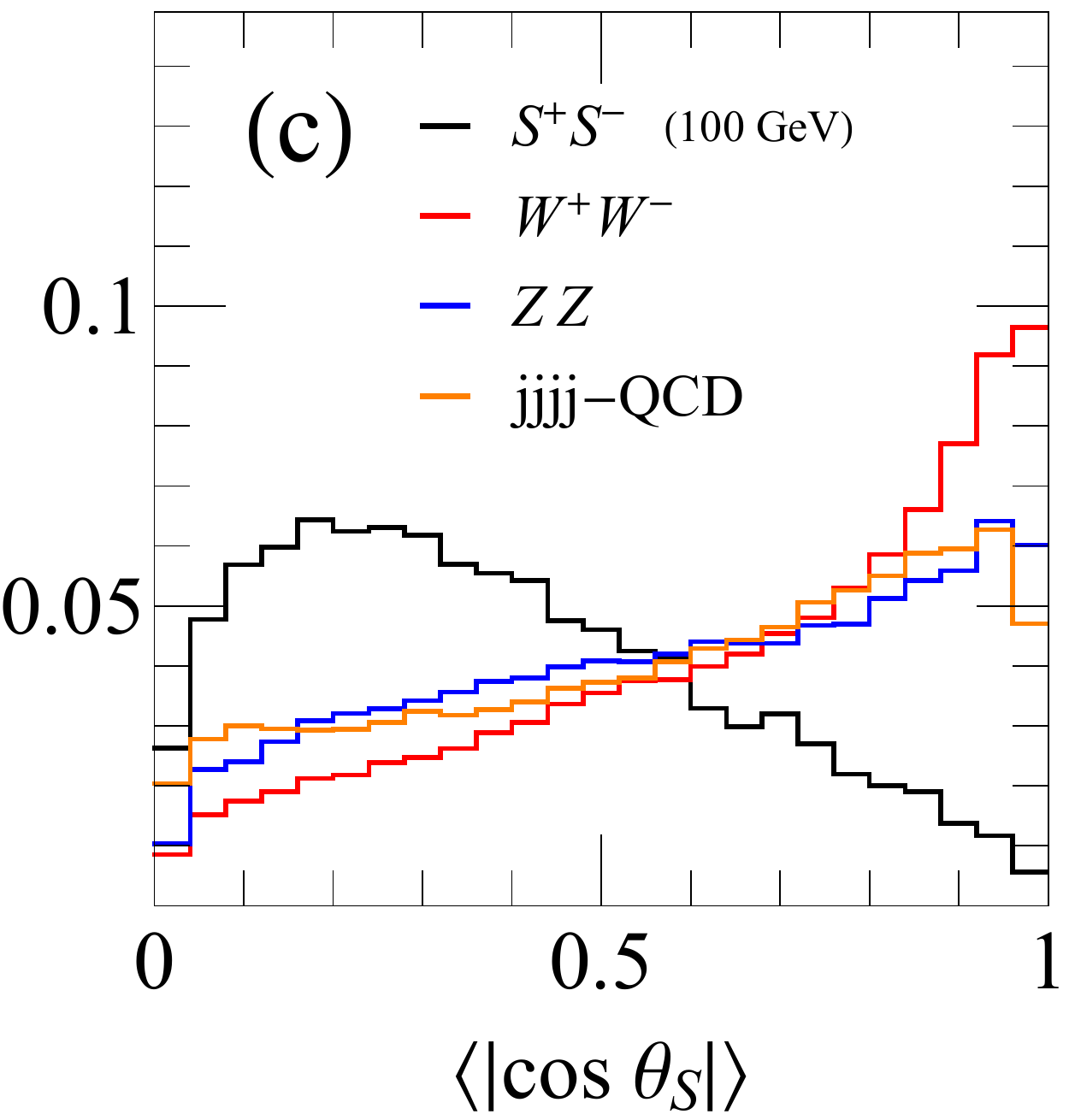}
\caption{The normalized distributions of (a) $p_T^{j_3}$, (b) $p_T^{j_4}$ and (c) $\langle |\cos\theta_S|\rangle$ in the $jj jj$ channel after the corresponding sequential cuts shown in Table~\ref{tb:jjjj_CEPC_tau_0p6}.}
\label{fg:jjjj}
\end{figure}

At last, we turn to the fully hadronic mode. In the selection cuts we demand at least four jets with $p_T > 5~\mathrm{GeV}$  appear in the central region ($|\eta| < 3$) of the detector. We further veto charged leptons with $p_T>5~{\rm GeV}$ and $|\eta|<3$. The four jets are ordered in terms of $p_T$. The cross sections of the signal and the background processes after the selection cuts are given in Table~\ref{tb:jjjj_CEPC_tau_0p6}. In addition to the $W^+W^-$ process, we now have substantial background from the $jjjj$-QCD process. Since the jets produced in the QCD process tend to be softer, we impose hard cuts on the $p_T$'s of final state jets. Fig.~\ref{fg:jjjj}(a) and (b) displays the normalized distributions of $p_T^{j_3}$ and $p_T^{j_4}$, respectively, for both signal and background processes. Here, $p_T^{j_3}$ ($p_T^{j_4}$) is defined as the third (fourth) jet ordered by their $p_T$'s. We impose the hard $p_T$ cuts as follows:  
\begin{eqnarray}
p_T^{j_3} > 20~\mathrm{GeV},\quad p_T^{j_4} > 12~\mathrm{GeV}.
\end{eqnarray}
The cross sections of the signal and background processes after the $p_T^{j_3, j_4}$ cut is also shown in the fourth column of Table~\ref{tb:jjjj_CEPC_tau_0p6}.

\begin{table}[b]
\scriptsize
\centering
\caption{Cross sections (in the unit of femtobarn) of the signal and backgrounds in the $jj j j$ mode at the CEPC. The kinematic cuts listed in each column are applied sequentially.}
\label{tb:jjjj_CEPC_tau_0p6}
\begin{tabular}{c | c | c | c | c}
\hline
$jj j j$ & No cut & Selection & $p_T^{j_3, j_4}$ cut & $\langle |\cos\theta_S|\rangle$ cut \\
\hline
$m_S = 100~\mathrm{GeV}$ &  \multirow{2}{*}{90.3}	&	 \multirow{2}{*}{88.0}	&	 \multirow{2}{*}{66.2}	&	 \multirow{2}{*}{51.3} \\
$\mathcal{B}_j = 1$ & & & & \\
\hline
\hline
$W^+W^-$ & 16520	&	7202	&	4367	&	1668\\
\hline
$ZZ$ & 1100	&	520	&	344	&	164\\
\hline
$Zh$ & 212	&	127	&	87.9	&	61.6\\
\hline
$W^\pm q\bar q^\prime$ & 307	&	194	&	131	&	75.4\\
\hline
$Zq\bar q$ & 418	&	214	&	129	&	82.8\\
\hline
$jjjj$-QCD & 15280	&	2810	&	627	&	370\\
\hline
\end{tabular}
\end{table}

Thanks to the absence of large missing energy in this channel, we can reconstruct the kinematics of the intermediate particles directly from the four jets in final state. To remove the ambiguities in the jet combination, we define the correct combination as the one that yields the least mass difference between any two jet pairs. After finding the correct jet pairs, we reconstruct the charged scalars in the intermediate state.  Figure~\ref{fg:jjjj}(c) displays the normalized distributions of $\langle |\cos\theta_S|\rangle$, which is defined as the average value of $|\cos\theta_S|$ of the two reconstructed charged scalar $S^\pm$'s. Here, $\theta_S$ is the polar angle of $S^\pm$ in the lab frame. A cut of 
\beq
\langle |\cos\theta_S|\rangle < 0.6,
\eeq
efficiently reduces the dominant $W^+W^-$ background, as well as other backgrounds, but retain the most of the signal events.  See the fifth column of Table~\ref{tb:jjjj_CEPC_tau_0p6}. We need an integrated luminosity of around $25~\mathrm{fb}^{-1}$ to reach a $5\sigma$ discovery for the signal process with $m_S = 100~\mathrm{GeV}$ and $\mathcal{B}_j =1$. 

After discovering the scalar $S^\pm$ in the $jjjj$ mode, the precision measurement on the mass and spin of $S^\pm$ is quite straightforward. Since we can fully reconstruct the event kinematics from the final state jets, the mass and spin determinations are very similar to those in the previous $\tau^\pm\nu jj$ mode. 

\subsection{Combined analysis}

Having discussed the cuts used in the analysis, we now present the discovery and exclusion limits on the singlet charged scalar $S^\pm$ at the CEPC. 
We divide our search channels into two groups, one consists of the $e^\pm\mu^\mp\nu\bar\nu + e^+ e^- \nu\bar\nu + \mu^+\mu^-\nu\bar\nu$ modes, the other contains the $\tau^+\tau^-\nu\bar\nu + \tau^\pm\nu jj + jjjj$ modes. The former combination is able to constrain the decay branching ratios of $\mathcal{B}_e$ and $\mathcal{B}_\mu$, while the latter one is sensitive to $\mathcal{B}_\tau$  and $\mathcal{B}_j$. Within each combination, say, $e^\pm\mu^\mp\nu\bar\nu + e^+ e^- \nu\bar\nu + \mu^+\mu^-\nu\bar\nu$, we first vary the ratio between $\mathcal{B}_e$ and $\mathcal{B}_\mu$, and then obtain the minimally reachable value of $\mathcal{B}_e + \mathcal{B}_\mu$, denoted as $(\mathcal{B}_e + \mathcal{B}_\mu)_{\mathrm{min}}$,  for each assigned ratio. Finally, we retain the \emph{largest} $(\mathcal{B}_e + \mathcal{B}_\mu)_{\mathrm{min}}$ among all possible ratios  as the most \emph{conservative} lower bound on $\mathcal{B}_e + \mathcal{B}_\mu$. Similar treatment is also applied to the $\tau^+\tau^-\nu\bar\nu + \tau^\pm\nu jj + jjjj$ combination, where we instead look for the smallest $(\mathcal{B}_e + \mathcal{B}_\mu)_{\mathrm{max}}$ among all possible ratios between $\mathcal{B}_\tau$ and $\mathcal{B}_j$. 

\begin{figure}
\centering
\includegraphics[scale=0.5]{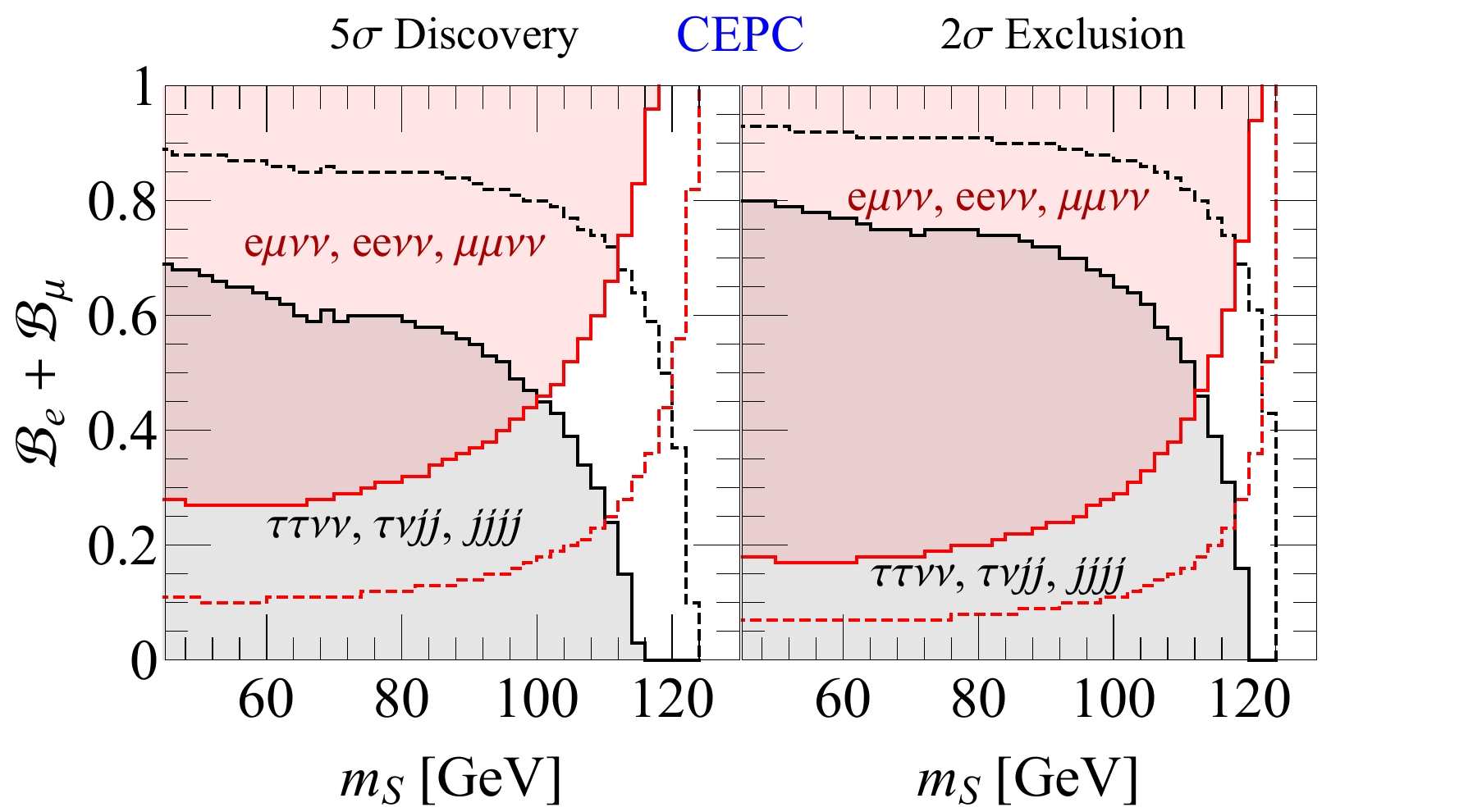}
\caption{The $5\sigma$ discovery parameter space (left) and the $2\sigma$ exclusion parameter space (right) at the CEPC. Red (gray) shaded regions are the reachable parameter space when combining $e^\pm\mu^\mp\nu\bar\nu, e^+ e^- \nu\bar\nu$ and $\mu^+\mu^-\nu\bar\nu$ ($\tau^+\tau^-\nu\bar\nu, \tau^\pm\nu jj$ and $jjjj$) channels with $ \mathcal{L} = 100~\mathrm{fb}^{-1}$. Dashed lines are the results for $ \mathcal{L} = 5~\mathrm{ab}^{-1}$.}
\label{fg:combined_CEPC}
\end{figure}

The obtained conservative lower/upper bounds on $\mathcal{B}_e + \mathcal{B}_\mu$ are then shown in Fig.~\ref{fg:combined_CEPC}, where left and right plots are for $5\sigma$ discovery and $2 \sigma$ exclusion at the CEPC, respectively. Red (gray) shaded regions are the reachable parameter space when combining $e^\pm\mu^\mp\nu\bar\nu$, $e^+ e^- \nu\bar\nu$ and $\mu^+\mu^-\nu\bar\nu$ ($\tau^+\tau^-\nu\bar\nu$, $\tau^\pm\nu jj$ and $jjjj$) channels with $ \mathcal{L} = 100~\mathrm{fb}^{-1}$. Dashed lines are the results for $ \mathcal{L} = 5~\mathrm{ab}^{-1}$. With $\mathcal{L} = 100~\mathrm{fb}^{-1}~(5~\mathrm{ab}^{-1})$ one is able to discover the singlet charged scalar up to $95~(118)~\mathrm{GeV}$, therefore, the unconstrained parameter space at the LEP, see Fig.~2 in Ref.~\cite{Cao:2017ffm}, can be completely covered by the CEPC, even at an early stage of running. In terms of exclusion, a mass of $m_S = 112~(122)~\mathrm{GeV}$ can be reached with $\mathcal{L} = 100~\mathrm{fb}^{-1}~(5~\mathrm{ab}^{-1})$. 

\section{Searching for $S^\pm$ at the ILC-350 and ILC-500}
\label{sec:ilc}

We now focus on the other two alternative scenarios of future lepton colliders, ILC-350 and ILC-500. We will not repeat the details of collider simulation for these two scenarios, as they are almost the same as the previous CEPC case.  The applied cuts are also similar, although in a few cases different cuts are used and the specific values of some cuts are adjusted due to higher center of mass energies. The finally obtained cut flow tables for six different search channels are given in Appendix~\ref{sec:ILC_cut_flow_tables}. 

In Fig.~\ref{fg:combined_ILC_350_500} we show the $5\sigma$ discovery potential (left) and $2\sigma$ exclusion limit (right) at ILC-350 (top) and ILC-500 (bottom). As in the previous CEPC case, we here also present the most conservative upper/lower bound on $\mathcal{B}_e + \mathcal{B}_\mu$ from two different combinations of search channels. Descriptions of lines and regions are the same as the previous case, except that the dashed lines are now for $\mathcal{L} = 1~\mathrm{ab}^{-1}$. Bumps of lines are due to the applied cuts and see Appendix~\ref{sec:ILC_cut_flow_tables} for detailed explanations. 

From Fig.~\ref{fg:combined_ILC_350_500} we observe that, if the scalar $S^\pm$ were indeed hiding in the unconstrained parameter space from the LEP below $80~\mathrm{GeV}$ (see Fig. 2 in Ref.~\cite{Cao:2017ffm}), a luminosity of $100~\mathrm{fb}^{-1}$ at both the ILC-350 and the ILC-500 would not be enough to discover it. However, with a higher luminosity of  $\mathcal{L} = 1~\mathrm{ab}^{-1}$ in both cases we can discover the singlet charged scalar up to $145~\mathrm{GeV}$ and $170~\mathrm{GeV}$, respectively. As for the exclusion limits, with $\mathcal{L} = 100~\mathrm{fb}^{-1}$ the ILC-350 (ILC-500) is able to exclude $m_S$ up to $140~\mathrm{GeV}$ ($150~\mathrm{GeV}$), respectively. At the ILC-350 such a limit can be further improved to $160~\mathrm{GeV}$ with $\mathcal{L} = 1~\mathrm{ab}^{-1}$, while all the mass region below $\sim m_t$ can be excluded at the ILC-500 with $\mathcal{L} = 1~\mathrm{ab}^{-1}$. 

\begin{figure}
\centering
\includegraphics[scale=0.5]{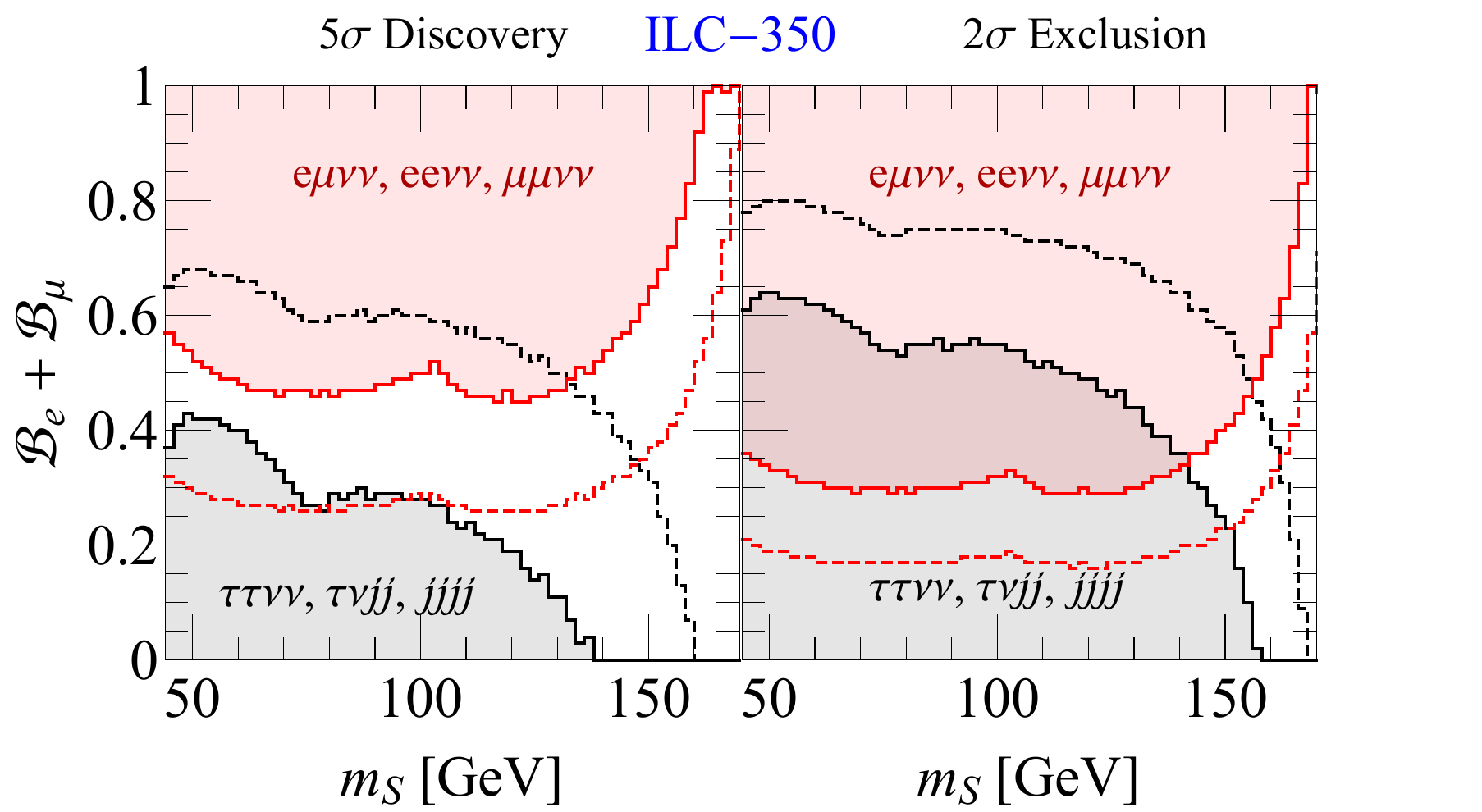}\\
\vskip 0.2cm
\includegraphics[scale=0.5]{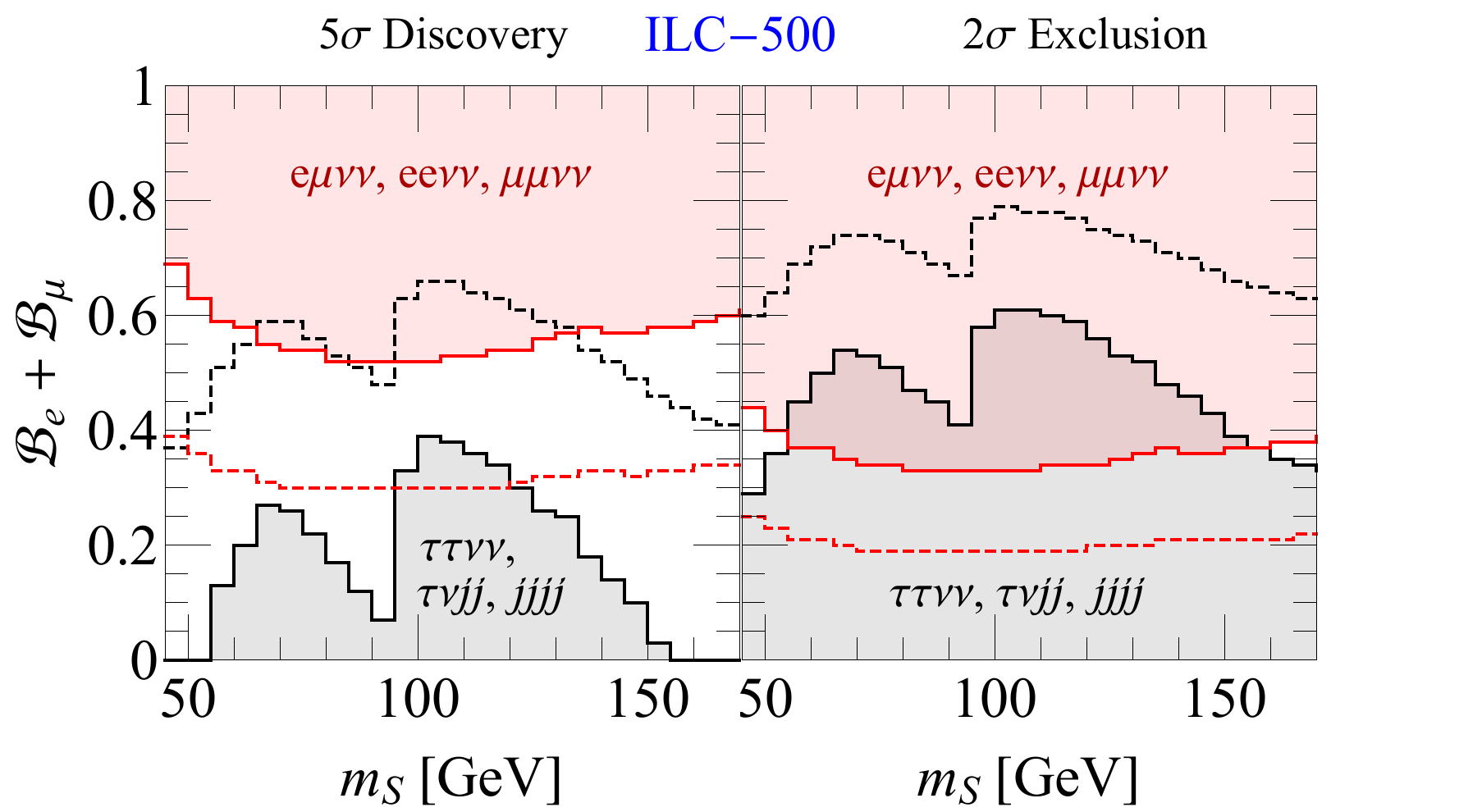}
\caption{$5\sigma$ discovery (left) and $2\sigma$ exclusion (right) plots at the ILC-350 (top) and the ILC-500 (bottom). Other descriptions are the same as Fig.~\ref{fg:combined_CEPC}, except that the dashed lines are for $\mathcal{L} = 1~\mathrm{ab}^{-1}$.}
\label{fg:combined_ILC_350_500}
\end{figure}

\section{Summary} \label{sec:summary}

Scalar sector of our Nature might be far more rich and complex than that of the SM, and the recent discovered Higgs boson can be just the tip of the iceberg. In this work we consider the possibility of a light weak singlet charged scalar $S^{\pm}$ with mass around $\mathcal{O}(100)~\mathrm{GeV}$, whose presence is still allowed by current experimental data. 

To describe its interactions with the SM particles, we adopt the effective field theory to write down gauge invariant operators involving $S^{\pm}$ up to dimension-5.  At the renormalizable level there exist only one operator with the singlet charged scalar and the SM fermions,  $f_{\alpha\beta} \overline{\ell}_{\mathrm{L}\alpha} \ell_{\mathrm{L}\beta}^c S$, whose coupling $f_{\alpha\beta}$ is suppressed by current limits from charged lepton flavor violation. We then move on to dimension-5, where four independent operators, i.e., $\bar e_{\rm R} e^c_{\rm R} SS$, $ \overline{Q}_{\rm L} H u_{\rm R}S$, $\overline{Q}_{\rm L}\widetilde{H} d_{\rm R} S^\dagger$ and $ \overline{\ell}_{\rm L} \widetilde{H} e_{\rm R} S^\dagger$, are identified after a careful treatment of field redefinition and gauge-fixing~\cite{Cao:2017ffm}. The finally obtained dominant decay modes of $S^-$ are then $S^- \rightarrow e^- \nu,~ \mu^- \nu,~ \tau^- \nu$ and $q \bar{q}^\prime$.

In this work, we focus on the discovery prospects and exclusion limits of searching for the singlet scalar $S^\pm$ at future lepton colliders. Three future lepton colliders are considered: CEPC ($\sqrt{s} = 250~\mathrm{GeV}$), ILC-350 and ILC-500. We show that, no matter how the charged scalar decays, a scalar with a mass up to 112 GeV can be discovered at the CEPC with $\mathcal{L} = 5~\mathrm{ab}^{-1}$ after combining all the possible decay models, while the charged scalar with the mass below 145 GeV (170 GeV) can certainly be discovered at the ILC-350 (ILC-500) with $\mathcal{L} = 1~\mathrm{ab}^{-1}$, respectively. 

\begin{acknowledgments}
We thank Ran Ding, Yandong Liu, Lian-Tao Wang and Bin Yan for useful discussions. The work is supported in part by the National Science Foundation of China under Grant Nos. 11275009, 11675002, 11635001 and 11725520, and by the China Postdoctoral Science Foundation under Grant No. 2017M610008. KPX is supported by the National Research Foundation of Korea under grant 2017R1D1A1B03030820. GL is supported by the Ministry of Science and Technology (MOST) under Grant No. 106-2112-M-002-003-MY3.
\end{acknowledgments}

\appendix

\section{Cut Flow Tables for the ILC-350 and ILC-500} \label{sec:ILC_cut_flow_tables}

In the appendix we present the cut flow tables for the ILC-350 (Tables \ref{tb:dilepton_ILC_350} and \ref{tb:ta_jj_ILC_350_tau_0p6}) and the ILC-500 (Tables \ref{tb:dilepton_ILC_500} and \ref{tb:ta_jj_ILC_500_tau_0p6}). In both cases the basic selection cuts are chosen the same as those used at the CEPC, while in the subsequent analysis we adjust the cut values in accord to the higher center of mass energies, and also introduce new or different cuts so as to more efficiently reduce background. 

In the analysis of the dilepton channel we add a hard cut of $M_{T2} > 80~\mathrm{GeV}$ to enhance the sensitivity for $m_S \gtrsim 100~\mathrm{GeV}$. Hence, the cuts used in the dilepton channel are summarized as follows: 
\bea
\text{Cut-1}~&:& ~ M_{T2} > 30 ~\text{GeV}, \nn\\
\text{Cut-2}~&:& ~\cos\theta^{\ell^\pm} \lessgtr \pm 0.3, \nn\\
\text{Cut-3}~&:& ~M_{T2} > 80 ~\text{GeV}. 
\label{A1}
\eea
In the analysis of the $\tau^{\pm}\nu j j$ channel at the ILC-500 we replace the cut on $\Phi_{\mathrm{acol}}^{jj}$ with a cut on $m_{jj}$, with $m_{jj}$ being the invariant mass of the dijet system, because the latter has a sharper drop at high energy end. The cuts used in the $\tau^+\tau^- \nu\bar{\nu}$ channel at the ILC-350 are given as follows:
\begin{align}
&\text{Cut-I}~:  &&E_\ell<45~\text{GeV}, \nn\\
&\text{Cut-II}~: &&M_{T2}>20~\text{GeV}, \nn\\
&\text{Cut-III}~: &&p_T^{\tau\text{-can1}}>40~\text{GeV}, 
\label{A2}
\end{align}
while at the ILC-500 are 
\begin{align}
&\text{Cut-I}~:  &&E_\ell<65~\text{GeV}, \nn\\
&\text{Cut-II}~: &&M_{T2}>30~\text{GeV}, \nn\\
&\text{Cut-III}~: &&p_T^{\tau\text{-can1}}>40~\text{GeV}. 
\label{A3}
\end{align}
The cuts used in the $\tau^{\pm}\nu jj$ channel at the ILC-350 are
\begin{align}
&\text{Cut-IV}~: &&\cancel{E}_T>20~\text{GeV}, \nn\\
&\text{Cut-V}~: &&E_{\text{bo}}^\tau<30~\text{GeV}, \nn\\
&\text{Cut-VI}~: &&Q_\tau\cdot\cos\theta_{\text{bo}}<0.5, \nn\\
&\text{Cut-VII}~: &&\Phi_{\text{acol}}^{jj}<1.2, 
\label{A4}
\end{align}
while at the ILC-500 are 
\begin{align}
&\text{Cut-IV}~: &\cancel{E}_T>35~\text{GeV}, \nn\\
&\text{Cut-V}~: &E_{\text{bo}}^\tau<30~\text{GeV}, \nn\\
&\text{Cut-VI}~: &Q_\tau\cdot\cos\theta_{\text{bo}}<0.5,\nn\\ 
&\text{Cut-VII}~: &m_{jj} > 90~\mathrm{GeV}. 
\label{A5}
\end{align}
The cuts used in the $jj jj$ channel at the ILC-350 are
\begin{align}
&\text{Cut-VIII}~: &p_T^{j_3(j_4)} > 20(12)~\text{GeV},\nn\\
&\text{Cut-IX}~: &\langle |\cos\theta^{\text{bo}}|\rangle < 0.6. 
\label{A6}
\end{align}
while at the ILC-500 are 
\begin{align}
&\text{Cut-VIII}~: &p_T^{j_3(j_4)} > 20(15)~\text{GeV},\nn\\
&\text{Cut-IX}~: &\langle |\cos\theta^{\text{bo}}|\rangle < 0.6. 
\label{A7}
\end{align}

The bumps observed in Fig.~\ref{fg:combined_ILC_350_500} can be understood from the cuts shown above. Although seemingly only in the $\tau^+\tau^-\nu\nu + \tau^\pm\nu j j + jjjj$ combined channel of ILC-500 are two prominent bumps observed, in fact they also exist for other cases. We take the combined channel of $\tau^+\tau^-\nu\nu+\tau^{\pm}\nu j j + jjjj$  at the ILC-500 as an example to discuss the origin of bumps. 

First of all, among the three channels $\tau^+\tau^-\nu\nu$, $ \tau^\pm\nu j j$ and $jjjj$ we identify that it is the $\tau^\pm\nu j j$ channel leads to the two bumps. The presence of the bump at a higher value of $m_S$ is due to the cut of $m_{jj} > 90~\mathrm{GeV}$, which is optimized for the case of $m_S \gtrsim 100~\mathrm{GeV}$. From the bottom plots of Fig.~\ref{fg:combined_ILC_350_500}, one can see that the sensitivity indeed becomes increasing when $m_S > 90~\mathrm{GeV}$, and its later drop is mainly due to the decrease of signal production cross section with the increase of $m_S$. On the other hand, the bump appearing at the lower energy end originates from the selection cuts, where we require one $\tau$ candidate and two jets. Were the mass of $S^{\pm}$ too small, the decay products would be very soft in the rest frame of $S^{\pm}$. In that case, only when the decay products are boosted by $S^{\pm}$ can they pass the selection cuts in the lab frame. However, since in the rest frame of $S^{\pm}$ its two-body decay products are always back-to-back, thus one is boosted to be hard while the other turns out to be so soft that it may not be tagged. Therefore, the requirements of our selection cuts, which include the full reconstruction of two jets, are hard to satisfy, leading to a reduced signal efficiency at a lower value of $m_S$. It thus explains the existence of the bump at the lower energy end. 

\begin{table*}
\renewcommand\arraystretch{1.4}

\begin{minipage}{.45\textwidth}
\caption{Cross sections (in the unit of femtobarn) of the signal and backgrounds in the dilepton channel at the ILC-350. The Cut-1, cut-2 and cut-3 are given in Eq.~\ref{A1}. The kinematic cuts listed in each column are applied sequentially.}
\label{tb:dilepton_ILC_350}
\begin{tabular}{c | c | c | c | c | c }
\hline
\hline
Dilepton, DF & No cut & Selection & Cut-1& Cut-2 & Cut-3 \\
\hline
$m_S = 100~\mathrm{GeV}$  & \multirow{2}{*}{65.3}	&	\multirow{2}{*}{24.4}	&	\multirow{2}{*}{19.3}	&	\multirow{2}{*}{10.1}	&	\multirow{2}{*}{2.0} \\
 $ \mathcal{B}_e = \mathcal{B}_\mu = 0.5$  & & & & & \\
 \hline
$m_S = 130~\mathrm{GeV}$ &  \multirow{2}{*}{35.6}	&	\multirow{2}{*}{13.5}	&	\multirow{2}{*}{11.5}	&	\multirow{2}{*}{5.4}	&	\multirow{2}{*}{2.3} \\
$ \mathcal{B}_e = \mathcal{B}_\mu = 0.5$  & & & & & \\
 \hline
$W^+W^-$ & 26363	&	529.2	&	393.0	&	31.6	&	0 \\
 \hline
$ZZ$ & 1161	&	0.6	&	0.2	&	0.2	&	0\\
 \hline
$W^\pm e^\mp \nu$ & 2660	&	122.6	&	99.2	&	4.6	&	1.6\\
 \hline
$W^\pm\mu^\mp \nu$ & 134	&	11.2	&	9.0	&	1.0	&	0.2\\
 \hline
$Z\ell^+\ell^-$ & 636	&	4.2	&	0	&	0	&	0\\
 \hline
$e^+e^-\ell^+\ell^-$ (VBS) & 9585	&	21.6 	&	1.8	&	0	&	0\\
 \hline
$e^+e^-\tau^+\tau^-$ (VBS) & 2832	&	2.6	&	0	&	0	&	0\\
 \hline
$\tau^+\tau^-$ & 3067	&	48.8	&	0	&	0	&	0\\
 \hline
\hline
\end{tabular}
\vskip 0.25 cm
\begin{tabular}{c | c | c | c | c | c }
\hline
\hline
Dilepton, SF-ee & No cut & Selection & Cut-1 & Cut-2 & Cut-3 \\
\hline
$m_S = 100~\mathrm{GeV}$  & \multirow{2}{*}{65.3}	&	\multirow{2}{*}{10.5}	&	\multirow{2}{*}{7.2}	&	\multirow{2}{*}{3.9}	&	\multirow{2}{*}{0.9} \\
$\mathcal{B}_e = \mathcal{B}_\mu = 0.5$ & & & & & \\
\hline
$m_S = 130~\mathrm{GeV}$& \multirow{2}{*}{35.6}	&	\multirow{2}{*}{5.6}	&	\multirow{2}{*}{4.3}	&	\multirow{2}{*}{2.1}	&	\multirow{2}{*}{1.0} \\
 $\mathcal{B}_e = \mathcal{B}_\mu = 0.5$ & & & & & \\
\hline
$W^+W^-$ & 26363	&	219.4	&	133.4	&	11.0	&	0 \\
\hline
$ZZ$ & 1161	&	0.3	&	0.1	&	0	&	0\\
\hline
$W^\pm e^\mp \nu$ & 2660	&	96.3	&	66.5	&	2.9	&	1.1\\
\hline
$Z\ell^+\ell^-$ & 636	&	14.6	&	7.9	&	0.4	&	0.2\\
\hline
$e^+e^-\ell^+\ell^-$ (VBS) & 9585	&	102.1	&	0.9	&	0.2	&	0\\
\hline
$e^+e^-\tau^+\tau^-$ (VBS) & 2832	&	2.1	&	0.2	&	0	&	0\\
\hline
$\tau^+\tau^-$ & 3067	&	19.0	&	0	&	0	&	0\\
\hline
\hline
\end{tabular}
\vskip 0.25 cm
\begin{tabular}{c | c | c | c | c | c }
\hline
\hline
Dilepton, SF-$\mu\mu$ & No cut & Selection & Cut-1 & Cut-2 & Cut-3 \\
\hline
$m_S = 100~\mathrm{GeV}$  & \multirow{2}{*}{65.3}	&	\multirow{2}{*}{11.6}	&	\multirow{2}{*}{8.0}	&	\multirow{2}{*}{4.3}	&	\multirow{2}{*}{0.9} \\
$\mathcal{B}_e = \mathcal{B}_\mu = 0.5$ & & & & & \\
\hline
$m_S = 130~\mathrm{GeV}$& \multirow{2}{*}{35.6}	&	\multirow{2}{*}{6.2}	&	\multirow{2}{*}{4.8}	&	\multirow{2}{*}{2.3}	&	\multirow{2}{*}{1.1} \\
 $\mathcal{B}_e = \mathcal{B}_\mu = 0.5$ & & & & & \\
 \hline
$W^+W^-$ &  26363	&	253.3	&	148.6	&	12.3	&	0\\
 \hline
$ZZ$ & 1161	&	0.3	&	0.2	&	0.1	&	0\\
 \hline
$W^\pm\mu^\mp \nu$ & 134	&	10.9	&	7.6	&	0.8	&	0.3\\
 \hline
$Z\ell^+\ell^-$ & 636	&	3.3	&	0.8	&	0.3	&	0.2\\
 \hline
$e^+e^-\ell^+\ell^-$ (VBS) & 9585	&	112.8	&	0	&	0	&	0\\
 \hline
$e^+e^-\tau^+\tau^-$ (VBS) & 2832	&	0.3	&	0	&	0	&	0\\
 \hline
$\tau^+\tau^-$ & 3067	&	22.6	&	0	&	0	&	0\\
 \hline
\hline
\end{tabular}
\end{minipage} \qquad
\begin{minipage}{.5\textwidth}
\caption{Cut flow tables  of the $\tau^+\tau^- \nu\bar{\nu}$, $\tau^{\pm}\nu jj$ and $jj jj$ channels at ILC-350. The Cut-I to Cut-IX are given in Eqs.~\ref{A2}, \ref{A4} and \ref{A6}. The kinematic cuts listed in each column are applied sequentially.}
\label{tb:ta_jj_ILC_350_tau_0p6}
\begin{tabular}{c | c | c | c | c | c }
\hline
\hline
$\tau^+\nu\tau^-\bar{\nu}$ & No cut & Selection & Cut I & Cut II & Cut III \\
\hline
$m_S = 100~\mathrm{GeV}$ & \multirow{2}{*}{65.3}	&	\multirow{2}{*}{21.2}	&	\multirow{2}{*}{15.8}	&	\multirow{2}{*}{8.7}	&	\multirow{2}{*}{6.1} \\
$\mathcal{B}_\tau  = 1$ & & & & & \\
\hline 
$W^+W^-$ & 26363	&	1464	&	203.4	&	131.4	&	34.4 \\
\hline 
$ZZ$ & 1161	&	26.7	&	3.5	&	2.3	&	1.5 \\
\hline 
$W^\pm e^\mp \nu$ & 2660	&	275.7	&	5.0	&	3.6	&	0.7 \\
\hline 
$W^\pm\mu^\mp \nu$ & 134	&	27.5	&	0.3	&	0.2	&	0.1 \\
\hline 
$W^\pm\tau^\mp\nu$ & 134	&	18.3	&	6.1	&	4.2	&	2.4 \\
\hline 
$Z\ell^+\ell^-$ & 636	&	25.5	&	1.6	&	0.3	&	0 \\
\hline 
$e^+e^-\ell^+\ell^-$ (VBS) & 9585	&	222	&	121	&	18.8	&	0.1 \\
\hline 
$e^+e^-\tau^+\tau^-$ (VBS) & 2832	&	42.7	&	29.8	&	0.7	&	0 \\
\hline 
$\tau^+\tau^-$ & 3067	&	701	&	422	&	0.5	&	0.5 \\
\hline 
$\gamma \gamma \rightarrow \tau^+\tau^-$ & \multirow{2}{*}{13406}	&	\multirow{2}{*}{24.4}	&	\multirow{2}{*}{24.4}	&	\multirow{2}{*}{0}	&	\multirow{2}{*}{0} \\
Beamstrahlung & & & & & \\
\hline 
\hline
\end{tabular}
\vskip 0.25 cm
\begin{tabular}{c | c | c | c | c | c | c  }
\hline
\hline
$\tau^\pm\nu j j $ & No cut & Selection & Cut IV & Cut V & Cut VI & Cut VII \\
\hline
$m_S = 100~\mathrm{GeV}$ & \multirow{2}{*}{65.3}	&	\multirow{2}{*}{12.3}	&	\multirow{2}{*}{12.0}	&	\multirow{2}{*}{6.6}	&	\multirow{2}{*}{5.7}	&	\multirow{2}{*}{4.1} \\
$\mathcal{B}_\tau  = \mathcal{B}_j = 0.5$ & & & & & \\
\hline
$W^+W^-$ & 26363	&	5382	&	4686	&	835	&	175	&	41.0 \\
\hline
$ZZ$ & 1161	&	13.6	&	7.0	&	3.1	&	2.0	&	0.9 \\
\hline
$Zh$ & 191	&	4.6	&	3.3	&	1.4	&	1.1	&	1.0 \\
\hline
$W^\pm e^\mp \nu$ & 2660	&	488	&	444	&	2.4	&	0.9	&	0.4 \\
\hline
$W^\pm\mu^\mp \nu$ & 134	&	50.8	&	47.3	&	1.2	&	0.4	&	0.1 \\
\hline
$W^\pm\tau^\mp\nu$ & 134	&	34.2	&	30.6	&	11.3	&	3.0	&	1.3 \\
\hline
$W^\pm q\bar q^\prime$ & 818	&	101.1	&	87.9	&	19.7	&	6.0	&	5.6 \\
\hline
$Z\ell^+\ell^-$ & 636	&	71.6	&	9.6	&	2.1	&	1.9	&	0.6 \\
\hline
$e^+e^-\tau^+\tau^-$ (VBS) & 2832	&	7.6	&	2.8	&	0.5	&	0.2	&	0.2 \\
\hline
$e^+e^-q q^\prime$ (VBS) & 5858	&	31.3	&	2.3	&	0	&	0	&	0 \\
\hline
$jjjj$-QCD & 15221	&	13.5	&	5.3	&	4.0	&	2.5	&	0.8 \\
\hline
\hline
\end{tabular}
\vskip 0.25 cm
\begin{tabular}{c | c | c | c | c  }
\hline
\hline
$jjjj$ & No cut & Selection & Cut VIII & Cut IX \\
\hline
$m_S = 100~\mathrm{GeV}$ & \multirow{2}{*}{65.3}	&	\multirow{2}{*}{61.1}	&	\multirow{2}{*}{48.6}	&	\multirow{2}{*}{38.0} \\
$\mathcal{B}_j = 1$ & & & & \\
\hline
$W^+W^-$ & 26363	&	10846	&	6938	&	1792 \\
\hline
$ZZ$ & 1161	&	530	&	375	&	125 \\
\hline
$Zh$ & 191	&	113	&	77.1	&	54.1 \\
\hline
$W^\pm q\bar q^\prime$ & 818	&	514	&	389	&	175 \\
\hline
$Zq\bar q$ & 393	&	208	&	163	&	90.1 \\
\hline
$jjjj$-QCD & 15221	&	3255	&	1075	&	571 \\
\hline
\hline
\end{tabular}
\end{minipage} 
\end{table*}

\begin{table*}
\renewcommand\arraystretch{1.4}
\begin{minipage}{.45\textwidth}
\caption{Cross sections (in the unit of femtobarn) of the signal and backgrounds in the dilepton channel at the ILC-500. The Cut-1, Cut-2 and Cut-3 are given in Eq.~\ref{A1}. The kinematic cuts listed in each column are applied sequentially.}
\label{tb:dilepton_ILC_500}
\centering
\begin{tabular}{c | c | c | c | c | c }
\hline
\hline
Dilepton, DF & No cut & Selection & Cut-1 & Cut-2 & Cut-3 \\
\hline
$m_S = 100~\mathrm{GeV}$  & \multirow{2}{*}{46.1}	&	\multirow{2}{*}{8.6}	&	\multirow{2}{*}{8.1}	&	\multirow{2}{*}{5.8}	&	\multirow{2}{*}{0.8} \\
 $ \mathcal{B}_e = \mathcal{B}_\mu = 0.5$  & & & & & \\
 \hline
$m_S = 130~\mathrm{GeV}$ &  \multirow{2}{*}{37.3}	&	\multirow{2}{*}{7.0}	&	\multirow{2}{*}{6.7}	&	\multirow{2}{*}{4.5}	&	\multirow{2}{*}{1.6} \\
$ \mathcal{B}_e = \mathcal{B}_\mu = 0.5$  & & & & & \\
 \hline
$W^+W^-$ & 16782	&	295.4	&	212.4	&	22.2	&	0 \\
 \hline
$ZZ$ & 706	&	0.6	&	0.2	&	0	&	0\\
 \hline
$W^\pm e^\mp \nu$ & 4045	&	172.2	&	146.4 	&	9.0	&	4.6\\
 \hline
$W^\pm\mu^\mp \nu$ & 1117	&	90.8 	&	73.6	&	10.2 	&	4.4\\
 \hline
$Z\ell^+\ell^-$ & 640	&	6.0		&	1.8	&	1.0	&	0\\
 \hline
$e^+e^-\ell^+\ell^-$ (VBS) & 5178	&	18.6	&	3.6	&	0	&	0\\
 \hline
$e^+e^-\tau^+\tau^-$ (VBS) & 1873	&	2.4	&	0.2	&	0	&	0\\
 \hline
$\tau^+\tau^-$ & 1537	&	26.8 	&	0	&	0	&	0\\
 \hline
\hline
\end{tabular}
\vskip 0.25 cm
\begin{tabular}{c | c | c | c | c | c }
\hline
\hline
Dilepton, SF-ee & No cut & Selection & Cut-1 & Cut-2& Cut-3\\
\hline
$m_S = 100~\mathrm{GeV}$  & \multirow{2}{*}{46.1}	&	\multirow{2}{*}{7.7}	&	\multirow{2}{*}{5.1}	&	\multirow{2}{*}{3.7}	&	\multirow{2}{*}{0.7} \\
$\mathcal{B}_e = \mathcal{B}_\mu = 0.5$ & & & & & \\
\hline
$m_S = 130~\mathrm{GeV}$& \multirow{2}{*}{37.3}	&	\multirow{2}{*}{5.2}	&	\multirow{2}{*}{3.8}	&	\multirow{2}{*}{2.6}	&	\multirow{2}{*}{1.2} \\
 $\mathcal{B}_e = \mathcal{B}_\mu = 0.5$ & & & & & \\
\hline
$W^+W^-$ & 16782	&	166.8	&	77.5	&	7.0	&	0 \\
\hline
$ZZ$ & 706	&	0.2	&	0.1	&	0	&	0\\
\hline
$W^\pm e^\mp \nu$ & 4045	&	141.7	&	104.5	&	6.2	&	3.5\\
\hline
$Z\ell^+\ell^-$ & 640	&	15.9	&	9.3	&	0.5	&	0.2\\
\hline
$e^+e^-\ell^+\ell^-$ (VBS) & 5178	&	109.6	&	8.7	&	2.9	&	0\\
\hline
$e^+e^-\tau^+\tau^-$ (VBS) & 1873	&	1.5	&	0.1	&	0	&	0\\
\hline
$\tau^+\tau^-$ & 1537	&	12.6	&	0	&	0	&	0\\
\hline
\hline
\end{tabular}
\vskip 0.25 cm
\begin{tabular}{c | c | c | c | c | c }
\hline
\hline
Dilepton, SF-$\mu\mu$ &  No cut & Selection & Cut-1 & Cut-2 & Cut-3 \\
\hline
$m_S = 100~\mathrm{GeV}$  & \multirow{2}{*}{46.1}	&	\multirow{2}{*}{8.7}	&	\multirow{2}{*}{5.7}	&	\multirow{2}{*}{4.2}	&	\multirow{2}{*}{0.8} \\
$\mathcal{B}_e = \mathcal{B}_\mu = 0.5$ & & & & & \\
\hline
$m_S = 130~\mathrm{GeV}$& \multirow{2}{*}{37.3}	&	\multirow{2}{*}{5.8}	&	\multirow{2}{*}{4.4}	&	\multirow{2}{*}{3.0}	&	\multirow{2}{*}{1.3} \\
 $\mathcal{B}_e = \mathcal{B}_\mu = 0.5$ & & & & & \\
 \hline
$W^+W^-$ &  16782	&	145.5	&	87.5	&	8.6	&	0\\
 \hline
$ZZ$ & 706	&	0.2	&	0.1	&	0	&	0\\
 \hline
$W^\pm\mu^\mp \nu$ & 1117	&	89.4	&	64.2	&	8.8	&	4.1\\
 \hline
$Z\ell^+\ell^-$ & 640	&	2.2	&	1.5	&	0.5	&	0.3\\
 \hline
$e^+e^-\ell^+\ell^-$ (VBS) & 5178	&	123.2	&	8.7	&	3.4	&	0\\
 \hline
$e^+e^-\tau^+\tau^-$ (VBS) & 1873	&	0.5	&	0	&	0	&	0\\
 \hline
$\tau^+\tau^-$ & 1537	&	13.0	&	0	&	0	&	0\\
 \hline
\hline
\end{tabular}
\end{minipage}\quad
\begin{minipage}{.5\textwidth}
\caption{Cut flow tables  of the $\tau^+\tau^- \nu\bar{\nu}$, $\tau^{\pm}\nu jj$ and $jj jj$ channels at ILC-500. The Cut-I to Cut-IX are given in Eqs.~\ref{A3}, \ref{A5} and \ref{A7}. 
The kinematic cuts listed in each column are applied sequentially.}
\label{tb:ta_jj_ILC_500_tau_0p6}
\begin{tabular}{c | c | c | c | c | c }
\hline
\hline
$\tau^+\nu\tau^-\bar{\nu}$ & No cut& Selection & Cut I & Cut II & Cut III \\
\hline
$m_S = 100~\mathrm{GeV}$ & \multirow{2}{*}{46.1}	&	\multirow{2}{*}{4.1}	&	\multirow{2}{*}{3.1}	&	\multirow{2}{*}{0.9}	&	\multirow{2}{*}{0.8} \\
$\mathcal{B}_\tau  = 1$ & & & & & \\
\hline 
$W^+W^-$ & 16782	&	830	&	132	&	56.5	&	20.7 \\
\hline 
$ZZ$ & 706	&	15.3	&	2.1	&	1.2	&	1.0 \\
\hline 
$Zh$ & 85	&	0.7	&	0.4	&	0.4	&	0.3 \\
\hline 
$W^\pm e^\mp \nu$ & 4045	&	388	&	7.9	&	4.0	&	1.7 \\
\hline 
$W^\pm\mu^\mp \nu$ & 1117	&	218	&	2.6	&	1.8	&	1.2 \\
\hline 
$W^\pm\tau^\mp\nu$ & 1117	&	150	&	53.3	&	29.	&	22.3\\
\hline 
$Z\ell^+\ell^-$ & 640	&	25.0	&	2.9	&	0.5	&	0.1 \\
\hline 
$Z\nu\nu$ & 764	&	45.0	&	12.2	&	5.2	&	3.9\\
\hline 
$e^+e^-\ell^+\ell^-$ (VBS) & 5178	&	243	&	168	&	12.7	&	0.8 \\
\hline 
$e^+e^-\tau^+\tau^-$ (VBS) & 1873	&	41.3	&	34.5	&	0.3	&	0 \\
\hline 
$\tau^+\tau^-$ & 3067	&	341	&	208	&	0.1	&	0.1 \\
\hline 
$\gamma \gamma \rightarrow \tau^+\tau^-$ & \multirow{2}{*}{26007}	&	 \multirow{2}{*}{14.6}	&	 \multirow{2}{*}{14.6}	&	 \multirow{2}{*}{0}	&	 \multirow{2}{*}{0} \\
Beamstrahlung & & & & & \\
\hline 
\hline
\end{tabular}
\vskip 0.25 cm
\begin{tabular}{c | c | c | c | c | c | c  }
\hline
\hline
$\tau^\pm\nu j j $ & No cut & Selection & Cut IV & Cut V & Cut VI & Cut VII \\
\hline
$m_S = 100~\mathrm{GeV}$ & \multirow{2}{*}{46.1}	&	 \multirow{2}{*}{10.6}	&	 \multirow{2}{*}{10.2}	&	 \multirow{2}{*}{5.8}	&	 \multirow{2}{*}{5.0}	&	 \multirow{2}{*}{3.3}\\
$\mathcal{B}_\tau  = \mathcal{B}_j = 0.5$ & & & & & \\
\hline
$W^+W^-$ & 16782	&	3208	&	2185	&	413	&	89.9	&	0.2 \\
\hline
$ZZ$ & 706	&	11.6	&	3.2	&	1.6	&	1.0	&	0.3 \\
\hline
$Zh$ & 85	&	1.8	&	1.5	&	0.4	&	0.3	&	0.2 \\
\hline
$W^\pm e^\mp \nu$ & 4045	&	711	&	598	&	1.2	&	0.5	&	0 \\
\hline
$W^\pm\mu^\mp \nu$ & 1117	&	416	&	352	&	5.3	&	2.1	&	0\\
\hline
$W^\pm\tau^\mp\nu$ & 1117	&	287	&	230	&	74.2	&	15.9	&	0 \\
\hline
$W^\pm q\bar q^\prime$ & 687	&	77.1	&	52.9	&	14.3	&	3.6	&	3.4 \\
\hline
$Z\ell^+\ell^-$ & 640	&	95.4	&	3.8	&	0.1	&	0.1	&	0 \\
\hline
$e^+e^-\tau^+\tau^-$ (VBS) & 1873	&	4.1	&	0.9	&	0.1	&	0.1	&	0 \\
\hline
$e^+e^-q q^\prime$ (VBS) & 3203	 &	24.6	&	0.4	&	0	&	0	&	0 \\
\hline
$t\bar{t}$ & 891	&	11.9	&	10.0	&	2.1	&	1.7	&	1.3 \\
\hline
$jjjj$-QCD & 31521	&	29.3	&	9.3	&	5.4	&	3.1	&	1.0 \\
\hline
\hline
\end{tabular}
\vskip 0.25 cm
\begin{tabular}{c | c | c | c | c  }
\hline
\hline
$jjjj$ & No cut & Selection & Cut VIII & Cut IX \\
\hline
$m_S = 100~\mathrm{GeV}$ & \multirow{2}{*}{46.1}	&	\multirow{2}{*}{43.1}	&	\multirow{2}{*}{33.4}	&	\multirow{2}{*}{26.4} \\
$\mathcal{B}_j = 1$ & & & & \\
\hline
$W^+W^-$ & 16782	&	6049	&	3424	&	714 \\
\hline
$ZZ$ & 706	&	296	&	188	&	49.3 \\
\hline
$Zh$ & 85	&	49.5	&	37.3	&	27.6\\
\hline
$W^\pm q\bar q^\prime$ & 687	&	422	&	302	&	112 \\
\hline
$Zq\bar q$ & 230	&	122	&	87.8	&	48.9 \\
\hline
$t\bar{t}$ & 891	&	526	&	496	&	330 \\
\hline
$jjjj$-QCD & 31521	&	7976	&	2672	&	1330 \\
\hline
\hline
\end{tabular}
\end{minipage}
\end{table*}

\bibliographystyle{apsrev}
\bibliography{reference}

\end{document}